\documentclass[%
reprint, amsmath,amssymb, aps, superscriptaddress]{revtex4-2}
\usepackage{b26_style}

\newcommand{\remark}[1]{\textcolor{blue}{#1}}

\newcommand{\tint}{t_I}
\newcommand{\topt}{t^*}
\newcommand{\tRO}{t_R}

\def\suppl{1}
\DeclareUnicodeCharacter{2009}{\,}

\begin{document}

\title{Efficient entanglement of spin qubits mediated by a hot mechanical oscillator}
\author{Emma Rosenfeld}
\affiliation{Physics Department, Harvard University, Cambridge, MA 02318, USA.}
\email{erosenfeld@g.harvard.edu}
\author{Ralf Riedinger}
\affiliation{Physics Department, Harvard University, Cambridge, MA 02318, USA.}
\author{Jan Gieseler}\thanks{Current address: IAV GmbH DigitalLab}
\affiliation{Physics Department, Harvard University, Cambridge, MA 02318, USA.} 
\author{Martin Schuetz}\thanks{Previous address: Physics Department, Harvard University}
\affiliation{Amazon Quantum Solutions Lab, Seattle, Washington 98170, USA.}
\affiliation{AWS Center for Quantum Computing, Pasadena, California 91125, USA.}
\author{Mikhail D. Lukin}
\affiliation{Physics Department, Harvard University, Cambridge, MA 02318, USA.}

\date{\today}

\begin{abstract} 
Localized electronic and nuclear spin qubits in the solid state constitute a promising platform for storage and manipulation of quantum information, even at room temperature. 
However, the development of scalable systems requires the ability to entangle distant spins, which remains a challenge today. We propose and analyze an efficient, heralded scheme that employs a parity measurement in a decoherence free subspace to enable fast and robust entanglement generation between distant spin qubits mediated by a hot mechanical oscillator. We find that high-fidelity entanglement at cryogenic and even ambient temperatures is feasible with realistic parameters, and show that the entangled pair can be subsequently leveraged for deterministic controlled-NOT operations between nuclear spins.  Our results open the door for novel quantum processing architectures for a wide variety of solid-state spin qubits. 
\end{abstract}

\maketitle

\paragraph*{Introduction.}
Electronic and nuclear spin qubits in the solid state are encouraging candidates for the realization of quantum information systems. Over the past decade, long-lived quantum memories and few-qubit quantum registers have been demonstrated in several different platforms, including under ambient conditions. The key, outstanding challenge is associated with engineering fast, programmable interactions between spin qubits separated by micrometer-scale distances.
For example, color centers such as the nitrogen vacancy (NV) center in diamond are promising contenders as robust qubits, owing to their long coherence times at room temperature \cite{herbschleb_ultra-long_2019}, well-developed microwave control, and optical initialization and readout. However, generating entanglement on-demand between spins remains a challenge: the short-range dipole-dipole interaction limits connectivity \cite{dolde_room-temperature_2013}, while optical entanglement schemes are inefficient \cite{bernien_heralded_2013, humphreys_deterministic_2018, borregaard_heralded_2015}, require cryogenic temperatures, and induce decoherence on nuclear spin registers \cite{jiang_coherence_2008, kalb_dephasing_2018}. 

In a complementary approach, it was suggested \cite{rabl_quantum_2010} to transduce interactions via magnetically functionalized oscillators, leveraging recent advances in the control of micromechanical resonators \cite{tsaturyan_ultracoherent_2017, maccabe_phononic_2019}, which enables quantum control of solid-state electron spins \cite{rugar_single_2004, hong_coherent_2012, delord_spin-cooling_2020, lee_topical_2017, rao_heralded_2016}.  Robustness against thermal noise is desirable for such applications at elevated temperatures, to avoid phonon-induced gate errors \cite{sorensen_quantum_1999}. Previous approaches for such `hot' gates require a large qubit-resonator cooperativity $C \gg 1$ for low error rates \cite{schuetz_high-fidelity_2017}, with error scaling as $\mathcal{E} \propto 1/\sqrt{C}$ (the cooperativity $C \equiv \lambda^2 / \Gamma \kappa n_{th}$ compares the coherent coupling rate $\lambda$ to the dissipation rates of the spin and resonator, $\Gamma$ and $\kappa n_{th}$, respectively). These regimes are experimentally challenging to achieve, such that a demonstration of mechanically mediated entanglement remains elusive.

\begin{figure}
\begin{center}	
  \includegraphics[width=\columnwidth]{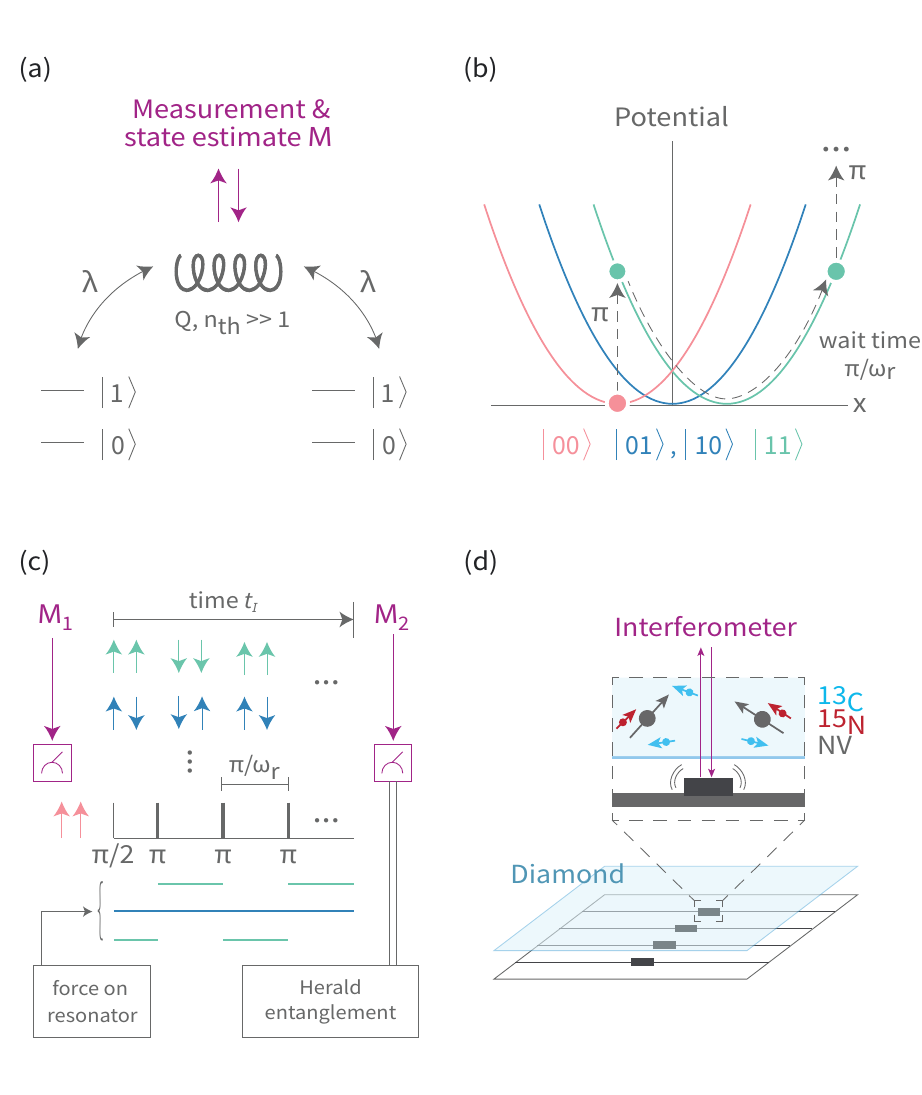}
  \caption{Entanglement protocol. \textbf{(a)} Two qubits are coupled with equal strength $\lambda$ to a hot resonator, which can be independently measured and that has high quality factor $Q$ and thermal occupation $n_{th}$. 
  \textbf{(b)} Spin-dependent potential of the resonator. Spin states $\ket{01}$, $\ket{10}$ are effectively decoupled, whereas $\ket{00}$, $\ket{11}$ shift the potential, such that toggling between $\ket{00}$ and $\ket{11}$ every half period drives the resonator.
  \textbf{(c)} The spins are initialized in $\ket{++}$ and the resonator states before ($M_1$) and after ($M_2$) applying pulse sequence (black) on the spins for duration $\tint$ are compared.
  Absence of displacement indicates spins are in the entangled anti-parallel states, as only parallel states (green) exert a force on the resonator.
  \textbf{(d)} Proposed implementation. 
  A diamond with NV centers is placed near a microresonator (grey) functionalized with nanomagnets (black), which is measured interferometrically.
  The entangled NV spin state is used to teleport gates between proximal $^{13}$C or $^{15}$N nuclear spins.}
  \label{fig:GeneralIdea}
\end{center}
\end{figure}
In this Letter, we propose and analyze a fast and robust entanglement protocol for two spin qubits (with eigenstates $\ket{0}$ and $\ket{1}$), linearly coupled to a common mode of a high-temperature mechanical resonator (Fig \ref{fig:GeneralIdea}(a)), via application of parity measurements in decoherence-free subspaces (DFS) \cite{hill_weak_2008, roch_observation_2014, martin_deterministic_2015}. In our approach, the  Bell states $\left| \Psi^\pm \right\rangle \propto \left|01 \right\rangle \pm \left|10 \right\rangle$ decouple from the resonator, forming a DFS insensitive to the thermal fluctuations of the hot resonator. In contrast, the aligned states $\left| 00 \right\rangle$ and $\left| 11 \right\rangle$ exert a force on the mechanical oscillator, resulting in a large displacement which can be observed (Fig \ref{fig:GeneralIdea}(b)). We can thus herald the entangled $\left| \Psi^\pm \right\rangle$ state by monitoring the absence of an excess force on the mechanical oscillator, constituting a (half) parity measurement in a measurement-free subspace \cite{hill_weak_2008, martin_what_2017, roch_observation_2014}. By design, this protocol is robust to thermal noise, and neither requires strong coupling, nor cooling to the mechanical quantum ground state. We show that entanglement can be generated at high success rates with relaxed cooperativity requirements, $C \gtrsim 1$, and with error scaling approaching $\mathcal{E} \propto \ln(C)/C$ at large cooperativity $C$. We specifically analyze an experimental realization involving NV centers in diamond, coupled to magnetically functionalized mechanical nano-beam resonators \cite{delord_spin-cooling_2020, hong_coherent_2012, kolkowitz_coherent_2012, rao_heralded_2016}, but note that it can equally be applied to other qubit species coupled to bosonic modes at high temperatures \cite{christle_isolated_2015, ali_momenzadeh_thin_2016, barfuss_strong_2015, barson_nanomechanical_2017, golter_optomechanical_2016, khanaliloo_single-crystal_2015, macquarrie_mechanical_2013, meesala_enhanced_2016, ovartchaiyapong_dynamic_2014, teissier_strain_2014, gieseler_single-spin_2020}, even in cases when high-fidelity readout is not available by other means \cite{rugar_single_2004, longenecker_high-gradient_2012, nichol_nanomechanical_2012, fischer_spin_2019}. 

The entangled pair of electronic spins can be subsequently leveraged to herald two-qubit gates between nearby, coherently-coupled nuclear spin memories in the solid state (Fig \ref{fig:GeneralIdea}(d)). Assuming state-of-the-art quality ($Q$) factors, spin-mechanics coupling strength, and spin coherence times \cite{tsaturyan_ultracoherent_2017, rossi_measurement-based_2018,abobeih_one-second_2018, guo_feedback_2019, arcizet_single_2011}, we expect that our entangling gate can achieve error rates below 1\% at cryogenic temperatures. With modest improvements in the coupling strength, similarly low error rates can be achieved at room temperature.

\paragraph*{Entanglement protocol.}
The key idea of our approach can be understood by considering two spin qubits, characterized by the Pauli operators $\sigma_{x,y,z}^{(i)}$ ($i=1,2$), that are linearly coupled with equal strength $\lambda$ to a micromechanical oscillator (Fig \ref{fig:GeneralIdea}(a), see [SI] for a treatment of inhomogenous $\lambda$). If the qubit frequencies $\omega_s^{(i)}$  ($i=1,2$) strongly exceed that of the resonator $\omega_r\ll \omega_s^{(i)}$, the transverse coupling terms can be ignored, and the system Hamiltonian is: 
\begin{equation}
    \mathcal{H}/\hbar = \frac{\omega_s^{(1)}}{2}\sigma_z^{(1)} + \frac{\omega_s^{(2)}}{2}\sigma_z^{(2)} +  \omega_r a^\dagger a + \lambda S_z (a + a^{\dagger}),
\end{equation}
where $S_z = \sigma_z^{(1)} + \sigma_z^{(2)}$ and $a$ ($a^{\dagger}$) are the bosonic annihilation (creation) operators of the resonator mode. For the two states $\ket{01}$ and $\ket{10}$, the qubits are decoupled from the resonator: the $S_z = 0$ states comprise a DFS, i.e. their phase is independent of the mechanical state. The other two states exert a force $\sim \pm 2\hbar \lambda / z_p$ on the resonator, where $z_p$ is the mechanical zero point fluctuation.

In our entanglement protocol (Fig \ref{fig:GeneralIdea}(c)), (i), the state of the resonator is first measured, while the two spins are initialized in the separable state $\ket{+}\otimes \ket{+}\propto \sqrt 2 \ket{\Psi^+}+\ket{11}+\ket{00}$. Then, (ii) the spins interact with the resonator for a time $\tint$, while being subjected to a special resonant decoupling sequence, such that the spin states $\ket{11}$ and $\ket{00}$ displace the resonator state. Finally (iii), the resonator displacement is measured. If the resulting displacement is below a threshold, the spins are projected into the Bell state $\ket{\Psi^+}$, indicating successful entanglement generation. The protocol can be made deterministic by repeating steps (i-iii) until success (about 2-3 repetitions in the regime of interest).

To estimate its practical performance, we assume the mechanical system can be described by a master equation, is weakly coupled ($Q\gg1$) to a hot thermal bath at rate $\kappa = \omega_r / Q$ and temperature $T\gg \hbar\omega_r / k_B$, and each qubit is dispersively coupled to an independent reservoir with dephasing rate $\Gamma$. The Gaussian state of the oscillator can be estimated interferometrically and independently of the spins. In practice, the effects of the measurement backaction and duration are negligible in near term realizations (see SI). To simplify the derivation, in the following we assume short backaction evading measurements of the mechanical quadrature \cite{vanner_pulsed_2011}, neglecting the measurement duration and avoiding a lower limit on the measurement uncertainty. 

\begin{figure}
	\includegraphics[width=\columnwidth]{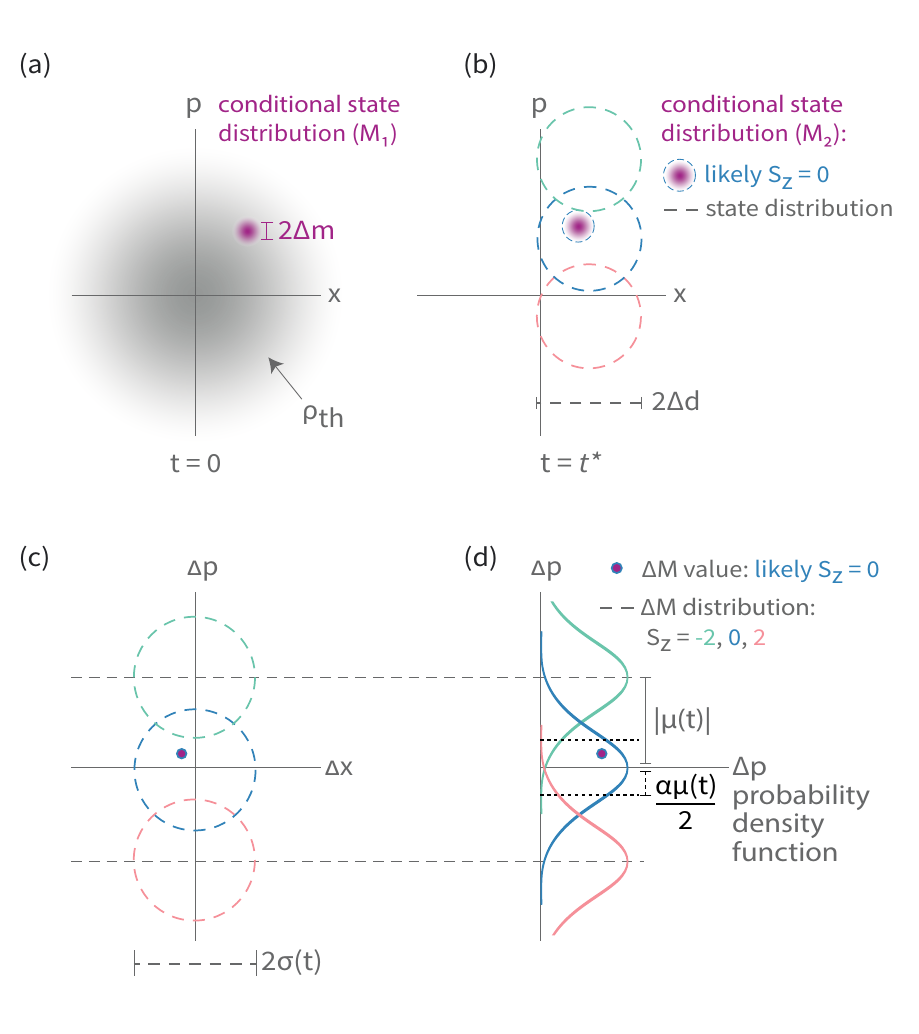}
	\caption{Mechanical phase space picture of the protocol. {\bf(a)} The thermal resonator ($\rho_{th}$) is localized by measurement $M_1$ (uncertainty $\Delta m$, purple) at time $t=0$ \cite{hofer_quantum_2017, hassani_further_2009, stengel_optimal_1994, bryson_applied_1975}. \textbf{(b)} After interaction time $\tint$, the spin-conditional resonator states (dashed circles, $S_z=2$ in pink, $S_z=0$ in blue, $S_z=-2$ in green) separate, and the resonator is measured again ($M_2$, purple). 
	\textbf{(c)} The conditional distribution of measured displacement (dashed circles) and \textbf{(d)} its projection onto the momentum basis is used to define a threshold $\alpha |\mu|/2$ (black dashed line for $\alpha \approx 0.6$). If a specific measurement (purple dot) lies within this threshold, $\langle S_z\rangle = 0$ is assigned and entanglement is heralded with fidelity $\mathcal{F}$.}
	\label{fig:PhaseSpace} 
\end{figure}

Figure \ref{fig:PhaseSpace} illustrates the key ingredients of the scheme. 
In step (i) we perform an initial linear measurement $M_1$ on the momentum quadrature $p \equiv i(a^\dagger-a)/\sqrt 2$, 
with measurement uncertainty $\Delta m$ 
(Fig \ref{fig:PhaseSpace}(a)). 
In step (ii), the spins are resonantly coupled to the oscillator by a series of $\pi$ pulses (here assumed to be ideal), with a pulse separation $2\tau=\pi/\omega_r$. This simultaneously maximizes the conditional mechanical displacement and the spin coherence by dynamically decoupling from their bath (Fig \ref{fig:GeneralIdea}(b) and \ref{fig:GeneralIdea}(c)). Throughout this pulse sequence, the force acting on the resonator is a square wave with amplitude $\sim -\hbar \lambda S_z / z_p $ (Fig \ref{fig:GeneralIdea}(c)) and frequency $\omega_r$. In the high $Q$ limit, higher harmonics of the force can be neglected, resulting in the effective interaction Hamiltonian in the rotating and toggling frame, under a rotating wave approximation [SI]
\begin{equation}\label{H_int}
    \mathcal{H}_{int}/\hbar = \frac{2\lambda}{\pi}S_z \left(a^{\dagger} + a\right),
\end{equation}
leading to a momentum shift of the resonator $\mu(S_z, \tint) = -4\sqrt{2}\lambda S_z(1-e^{-\kappa \tint/2})/\pi\kappa \approx -2\sqrt{2}\lambda S_z \tint/\pi$ in natural units, for $\kappa \tint \ll 1$. As the conditional equations of motion are linear, the motional states after step (ii) remain Gaussian with an uncertainty $\Delta d(\tint) \approx \sqrt{\kappa n_{th} \tint+\Delta m^2}$ (dashed circles in Fig. \ref{fig:PhaseSpace}(a)) and a spin-dependent expectation value of the momentum quadrature of $M_1 e^{-\kappa \tint/2} + \mu(S_z, \tint)$. Then, (iii) a second measurement $M_2$ localizes the resonator with uncertainty $\Delta m$, projecting the spin population to $\langle S_z \rangle\in\{0,+2,-2\}$ if the distributions are separable. If this is achieved within the coherence time $1/\Gamma$ of the spins, an $\langle S_z \rangle=0$ measurement projects the two spins into the entangled state $\ket{\Psi^+}$. 

A simple estimate shows the minimum requirements for this protocol. For negligible measurement uncertainties $\Delta m^2 \ll \kappa n_{th} \tint$, and an interaction time comparable to spin coherence time $\tint\sim 1/\Gamma$, yet shorter than the mechanical lifetime, $\tint\ll1/\kappa$, we find that the distributions become separable if the displacement $|\mu(2,1/\Gamma)|/2$ exceeds the uncertainty $\Delta d(1/\Gamma)$, i.e. $\lambda^2/\Gamma\kappa n_{th}=C\gtrsim 1$. 

To obtain an estimate of the fidelity, we compute the (Gaussian) probability density function $\mathcal{P}_{\langle S_z\rangle}(\Delta M)$ of the momentum difference $\Delta M \equiv M_2 - e^{-\kappa \tint/2}M_1$, conditional on the spin state $S_z$, which has expectation value $\mu(S_z, \tint)$ and variance
\begin{equation}\label{sigma_def}
    \sigma(\tint)^2 = \Delta m ^2(1+e^{-\kappa \tint}) + n_{th}(1-e^{-\kappa \tint}),
\end{equation}
corresponding to the contributions by the measurement uncertainties and diffusion during the interaction.
The state is assumed to have $S_z = 0$ if
\begin{equation}\label{threshold}
|\Delta M| < \alpha \mu(2, \tint)/2,
\end{equation}
where the threshold $\alpha \in (0, 1]$ can  be tuned to trade between a high acceptance rate ($\alpha\rightarrow 1$) and low false positive acceptance ($\alpha\rightarrow 0$). 
The probability of an accepted event being a true positive is $S(\alpha, g(\tint))\approx 1/(1 + e^{-2g(\tint)^2}) - \mathcal{O}(\alpha^2)$, given by the integrals of $\mathcal{P}_{0,\pm 1}$ within the thresholds, and weighted by the initial spin populations [SI], where we define the normalized displacement $g(\tint) \equiv \mu(2, \tint)/2\sigma(\tint)$. The error in entanglement fidelity $\mathcal{F} = \bra{\Psi^\pm} \rho \ket{\Psi^\pm}$ is the result of two independent error sources, namely spin dephasing and false positive $\langle S_z \rangle = 0$ assignments, yielding 
\begin{equation}\label{Fidelity}
    \mathcal{F} = \frac{1+e^{-2\Gamma \tint}}{2}S(\alpha, g(\tint)) \approx \frac{1}{2}\frac{1+e^{-2\Gamma \tint}}{1+e^{-2g(\tint)^2}} - \mathcal{O}(\alpha^2).
\end{equation}
It follows directly that entanglement can be generated ($\mathcal{F} > 1/2$) for $g(\tint)^2>\Gamma \tint$. We note that this simple estimate for $\alpha \rightarrow 0$ is a good approximation for general $\mathcal{F}$ [SI]. 
\paragraph*{Analysis of the Bell state preparation.}
\begin{figure}
	\includegraphics[width=\columnwidth]{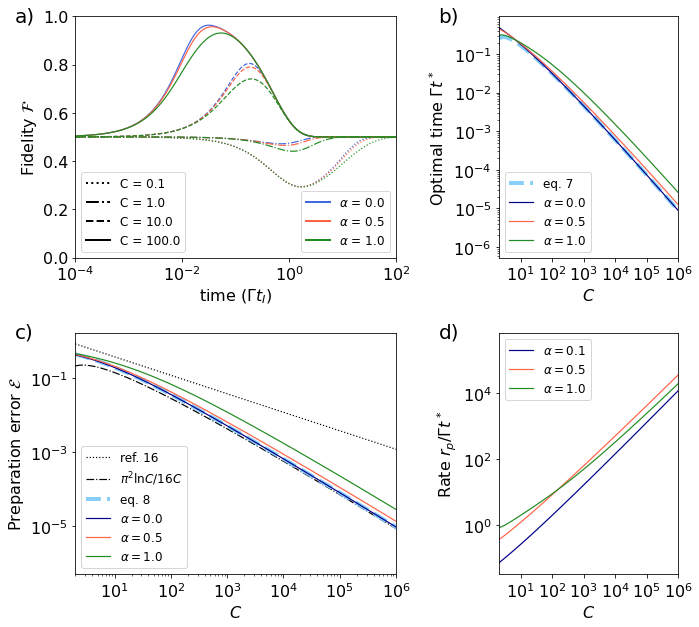}
	\caption{Performance. {\bf(a)} Fidelity $\mathcal{F}$ as a function of $\Gamma \tint$ under \eqref{gsq}. Different thresholds $\alpha$ are color coded and various cooperativities $C$ have the associated linestyles. The analytic form in eq. \eqref{Fidelity}, corresponding to $\alpha \ll 1$, is shown (blue curves). {\bf(b)} Optimal pulse sequence duration (solid lines, colors as in (a)), and the analytical approximation eq. \eqref{t_opt} (dashed light blue line). {\bf(c)} Infidelity $\mathcal{E}$ of the Bell state preparation (solid lines; colors as in (a)) and analytic approximation \eqref{Fidelity} for $\alpha \ll 1$ (dashed light blue line). The black dotted line represents the optimal infidelity of a deterministic hot gate \cite{schuetz_high-fidelity_2017}, and the dashed-dotted black line is the asymptotic $\pi^2 \text{ln}C/16C$ scaling.
	{\bf(d)} Normalized rate of true positive entanglement heralding events $r_p/\Gamma$. Colors as in (a), except $\alpha = 0.1$ (dark blue). The fast repetition rate allows multiple protocol attempts within the spin coherence time for large $C$.
	}
	\label{fig:FidelityAndTime} 
\end{figure}
In the following, we consider the experimentally relevant regime $\Delta m^2 \ll \kappa n_{th} \tint$, i.e.\ the linearized diffusion term dominates the variance of $\Delta M$, and that $\kappa \tint \ll 1$. In these limits
\begin{equation}\label{gsq}
g(\tint)^2 \approx (8/\pi^2) C \Gamma \tint,
\end{equation}
such that $\mathcal{F}$ can be described only in terms of $\alpha$, $C$ and $\Gamma \tint$ (Fig \ref{fig:FidelityAndTime}a).
The fidelity exceeds 1/2 for $C> \pi^2/8 \sim 1.2$, and exceeds 96\% for $C\sim100$, demonstrating that our protocol can be applied with relatively low $C$. The optimal interaction time $\Gamma \topt$ is determined numerically for each $C$ (Fig \ref{fig:FidelityAndTime}(b), dashed lines) and can be analytically approximated as
\begin{equation}\label{t_opt}
    \Gamma \topt \approx \frac{\pi^2}{16C}\text{ln}(16C/\pi^2-1)
\end{equation}
for $C> \pi^2/8$ and $\Delta m$, $\alpha\rightarrow0$ [SI], showing that the entanglement protocol is fast compared to the spin coherence time. 
We note that in the regime of interest ($C\gtrsim 8$, $\tint \ll 1/\Gamma$) decreasing the threshold $\alpha$ e.g.\ from 1 to 0.5 reduces the optimal interaction time (Fig. \ref{fig:FidelityAndTime}(b)). At high $C$, this can compensate the reduced acceptance rate for small $\alpha$ and increase the absolute rate of true positive entanglement heralding events, given by $r_p = \int_{-\theta}^{+\theta} \mathcal{P}_0(p)dp/2\topt$ for threshold $\theta = \alpha\mu(+2,\topt)/2$ (Fig. \ref{fig:FidelityAndTime}d) [SI]. Inserting equation \eqref{t_opt} into \eqref{Fidelity}, we find a lower bound to the fidelity
\begin{equation}\label{Error_scaling}
    \mathcal{F} \geq \frac{1}{2}\frac{1+(16C/\pi^2-1)^{-\frac{\pi^2}{8C}}}{1+(16C/\pi^2-1)^{-1}}
\end{equation}
again for $C> \pi^2/8$ and $\Delta m$,  $\alpha\rightarrow0$. 
The error $\mathcal{E} = 1-\mathcal{F}$ is shown in Fig \ref{fig:FidelityAndTime}(c). Remarkably, the cooperativity required to achieve an error  $\mathcal{E} <10^{-3}$ is more than two orders of magnitude lower than for previous mechanically mediated gates \cite{rabl_quantum_2010,schuetz_high-fidelity_2017}. For large $C$, $\mathcal{E} \sim (\pi^2/16) \ln(C)/C$. 
\paragraph*{Potential applications.}
The entanglement protocol presented here is inherently probabilistic, approaching a heralding probability of $1/2$ for $\alpha \rightarrow 1$. However, it can be extended to yield deterministic controlled-NOT gates between associated qubit registers by feedback, where here we assume a simple scheme, namely repeat until success \cite{humphreys_deterministic_2018}. In the following, we consider two electronic spins, such as NV centers (where the $\ket{m_s} = \ket{\pm 1}$ states are used as the two qubit states, for maximal displacements), interacting with the mechanical resonator and coupled to nearby $^{13}$C (or $^{15}$N) nuclear spins in the diamond host (see Fig. \ref{fig:GeneralIdea}d). The entangled NV spin state is used to teleport a gate between the nuclear spins \cite{chou_deterministic_2018}. 
Contributions to the gate error $\mathcal{E}_{T}$ include infidelities related to the ideal entanglement protocol $\mathcal{E}$, control ($\mathcal{E}_{C}$), initialization ($\mathcal{E}_{init}$) and readout ($\mathcal{E}_{RO}$) of the NV spins, as well as the electron-nuclear CNOT gate ($\mathcal{E}_{CNOT}$). Nuclear qubit errors arise from coupling to a bath ($\mathcal{E}_{nuc}$) at rate $\Gamma \gamma_N / \gamma_e$, where $\gamma_N$ ($\gamma_e$) is the nuclear (electron) gyromagnetic ratio, as well as dephasing due to electron spin control errors in failed entanglement attempts. As the latter depends on the heralding probability and the hyperfine coupling, we attribute it to $\mathcal{E}_C$ with a factor $\eta$, which is below 1 in the regime of interest [SI]. Combining state-of-the-art spin control \cite{rong_experimental_2015, harty_high-fidelity_2014} with robust decoupling sequences \cite{genov_arbitrarily_2017}, $\mathcal{E}_{C}$ can be neglected, however, if left unaddressed without optimal spin control, $\mathcal{E}_C$ can limit the fidelity \cite{kalb_dephasing_2018}. As the repetition rate is high (Fig. \ref{fig:FidelityAndTime}(b, d)), we further neglect the small probability of failure after a large number of repetitions in a synchronous circuit \cite{humphreys_deterministic_2018}. In this case, the total error of the deterministic nuclear gate is $\mathcal{E}_T = \mathcal{E} + 
2((1+\eta) \mathcal{E}_C +
\mathcal{E}_{init} + \mathcal{E}_{RO} +  \mathcal{E}_{CNOT} + \mathcal{E}_{nuc})$.
\begin{figure}
	\includegraphics[width=0.8\columnwidth]{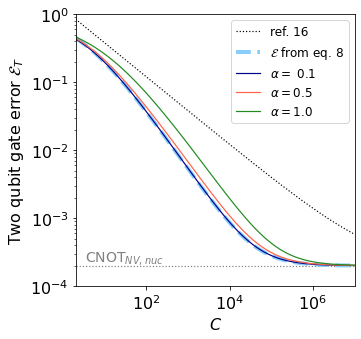}
	\caption{Error $\mathcal{E}_T$ for teleported CNOT gate between two $^{13}C$ (or $^{15}N$) nuclear spins, as a function of the cooperativity $C$ for various thresholds $\alpha$ and under \eqref{gsq}. A mechanical NV spin readout using an (arbitrary) interaction time of $20 \topt$ is assumed. The CNOT gate error between the NV and nuclear spin spin is set to $10^{-4}$ (dashed grey line) \cite{chou_optimal_2015, dong_precise_2020} and is also included in the representation of state-of-the-art hot gates \cite{schuetz_high-fidelity_2017}.
	Gate errors are plotted based on numerically calculated (solid lines) and analytically approximated $\mathcal{E}$ (eq. \eqref{Error_scaling}), neglecting nuclear spin decoherence, dashed light blue line). For details see text.}
	\label{fig:GateError} 
\end{figure}
\paragraph*{Experimental Implementation.}
In NV-based operations on nearby nuclear spins, optical excitation of the NV generally induces decoherence on the nuclear spins through the hyperfine coupling \cite{jiang_coherence_2008, reiserer_robust_2016, kalb_dephasing_2018, hopper_spin_2018}, which can limit $\mathcal{E}_T$ due to repeated spin initialization and readout. 
In our system, the mechanical oscillator can also be used for single-shot readout and initialization \cite{rugar_single_2004}, eliminating the need for optical illumination of the color center \cite{berman_stationary_2002}.

In practice, the measurement of the resonator could be implemented using an interferometer in conjunction with a Kalman filter (see e.g. Fig. \ref{fig:PhaseSpace}) \cite{wieczorek_optimal_2015, hassani_further_2009, hofer_quantum_2017, stengel_optimal_1994, bryson_applied_1975}, including spectator modes of the resonator as well as technical noise sources into the model [SI]. For further improvement, the estimation of the spin-induced displacement can be achieved with a multiple model adaptive estimation \cite{aguiar_convergence_2007, hanlon_multiple-model_2000}, while performing steps (ii) and (iii) simultaneously, as well as feedback on the spin state with additional global spin rotations during the pulse sequence, further increasing the entanglement fidelity and rate \cite{martin_deterministic_2015}. Finally, errors arising from small inhomoegenieties in the coupling strength can be suppressed with additional electron spin control [SI].

Fig. \ref{fig:GateError} shows the controlled-NOT gate error $\mathcal{E}_T$ as a function of $C$.  For realistic parameters described in [SI], at very high $C$, the total error $\mathcal{E}_T$ is limited by the electron-nuclear two qubit gate fidelity, while at more modest cooperativity, the error scales favorably compared to the existing state-of-the-art \cite{schuetz_high-fidelity_2017}. Note that an experimental demonstration of our protocol (with $\mathcal{E}_T \sim 10^{-1}$) may be possible at room temperature and $C \sim 8$, corresponding to state-of-the-art spin-mechanical systems ($1/\Gamma \sim 10$ms \cite{bar-gill_solid-state_2013}; $ Q \sim 10^9$ \cite{tsaturyan_ultracoherent_2017, ghadimi_elastic_2018}; $\lambda / 2\pi \sim 0.9$~kHz \cite{arcizet_single_2011}; $\Delta m^2 \sim 27$ \cite{guo_feedback_2019}). These parameters are within reach for a soft-clamped, silicon nitride nano-beam resonator, functionalized with a nano-magnet at the antinode of motion, and placed adjacent to diamond hosting NV centers (Fig. \ref{fig:GeneralIdea}(d)) \cite{ghadimi_elastic_2018, bar-gill_solid-state_2013}. Conversely, high fidelity gates ($\mathcal{E}_T < 10^{-2}$) at cryogenic temperature can be achieved with even smaller spin-mechanical coupling strength ($\lambda/2\pi \sim $100~Hz) and $1/\Gamma\sim 1$~s coherence time \cite{bar-gill_solid-state_2013, abobeih_atomic-scale_2019}, and at room temperature with modest improvement in spin-mechanics coupling strength ($\lambda \sim $2~kHz) [SI]. Such parameters yield high probability of success 
(approaching 50\% per run) and average gate duration approaching $10$ms, faster or comparable to deterministic protocols with the reported coupling strengths \cite{schuetz_high-fidelity_2017}. 
\paragraph*{Conclusion.}
We have proposed and analyzed a half-parity measurement protocol in a decoherence free subspace for entangling two qubits through a hot resonator, with error scaling that nears $\mathcal{E} \propto \ln(C)/C$. Our protocol is fast, robust to thermalization errors, and does not require ground state cooling. A teleported controlled-NOT gate employing the generated Bell pair is feasible with a solid-state system featuring magnetically functionalized resonators and solid state electronic spins with long coherence times \cite{ghadimi_elastic_2018, bar-gill_solid-state_2013}. 
While we analyzed an example implementation involving NV centers in diamond, as readout and initialization can be realized mechanically \cite{rugar_single_2004}, the protocol can also be applied to other promising paramagnetic defects, such as donor spins in silicon \cite{pla_single-atom_2012}. Further directions for analysis include leveraging continuous feedback to increase the rate of entanglement generation \cite{martin_what_2017, martin_deterministic_2015, martin_single-shot_2019}, as well as application of our protocol to generate multi-partite entangled states. Lastly, we note that for further improvements, nanobeam resonators can be electrostatically coupled \cite{rabl_quantum_2010}, using hybridized mechanical modes to selectively couple spins adjacent to distinct resonators, enabling multi-qubit connectivity far beyond reach of the magnetic dipole-dipole interactions alone. While our work leverages decades of development of micromechanical devices and solid state qubits, it simultaneously eliminates the need for high-fidelity single qubit optical or electronic addressing. With substantial technical improvement beyond the current state-of-the art, in the long term, this approach could potentially pave the way for realization of solid-state, room-temperature quantum information systems. 

\vspace{10pt}
\paragraph*{Acknowledgements} We would like to thank S. Hofer for insightful discussions. The authors thank C. Maxwell for design assistance with Figs. \ref{fig:GeneralIdea} and \ref{fig:PhaseSpace}. This work was supported by NSF, CUA, ARO MURI and V. Bush Faculty Fellowship. R.R. was supported by the Alexander von Humboldt Foundation.

\bibliographystyle{unsrt}
\bibliography{DFS_entangle} 

\if\suppl1

\clearpage


\onecolumngrid

\def\suppl{1}

\if\suppl0
\documentclass[a4paper,11pt]{article}
\usepackage[top=2.5cm, bottom=2cm, left=2.2cm, right=2.2cm]{geometry}

\usepackage{epsf,epsfig,changebar}
\usepackage{braket}
\usepackage{verbatim}
\usepackage{mathrsfs}
\usepackage[colorlinks]{hyperref}
\usepackage{fancyhdr}
\usepackage{empheq}
\usepackage[dvipsnames]{xcolor}
\usepackage{comment}
\usepackage{amsmath}
\usepackage{amssymb}
\usepackage{breqn}

\renewcommand{\baselinestretch}{1.02}
\newcommand{\floor}[1]{\lfloor #1 \rfloor}
\newcommand{\tint}{t_I}
\newcommand{\topt}{t^*}
\newcommand{\tRO}{t_R}
\pagestyle{fancy}
\renewcommand{\headrulewidth}{0pt}
\fancyhead{}
\fancyfoot{}
\chead{}
\rhead{\today}
\cfoot{\thepage}
\usepackage{color}
\usepackage{braket}
\usepackage{qcircuit}
\usepackage{booktabs}
\usepackage{subcaption}
\include{commands}
\newcommand{\remark}[1]{\textcolor{blue}{#1}}

\DeclareUnicodeCharacter{2009}{\,} 

\begin{document}
\fi

\begin{center}
{\Large
Efficient entanglement of spin qubits mediated by a hot mechanical oscillator \\
\vspace{3pt}
- Supplementary Information -}
\end{center}

\if\suppl0
\tableofcontents
\fi

\section{Effective Hamiltonian}

We consider two two-level systems, each coupled to a common bosonic mode described by annihilation operator $a$, which we assume here without loss of generality to be a mechanical resonator. In the case of equal coupling, the Hamiltonian reads
\begin{equation}\label{init_ham}
    \mathcal{H}/\hbar = \frac{\omega_s^{(1)}}{2}\sigma_z^{(1)} + \frac{\omega_s^{(2)}}{2}\sigma_z^{(2)} +  \omega_r a^\dagger a + \lambda S_z (a + a^{\dagger}),
\end{equation}
where $S_z = \sigma_z^{(1)} + \sigma_z^{(2)}$, and $\sigma_z^{(i)}$ is the Pauli $z$ matrix for qubit $i$. We assume $\lambda \ll \omega_r \ll  \omega_s$, and neglect terms that are $\sim \sigma_{x, y}^{(i)} \otimes (a + a^\dagger)$ in a secular approximation. Now consider adding a resonant, strong microwave drive on the spins such that the Rabi frequency $\Omega \gg \lambda$. If we exert a sequence of $\pi$ pulses such that the time $2\tau$ between the $\pi$ pulses is $\pi/\omega_r$, then each spin $\sigma_z^{(i)}(t) \to \sigma_z^{(i)}(0) \text{sgn}(\text{sin}(\omega_r t + \phi))$ (in the so-called `toggling frame'), where $\phi$ is set by the time between $t=0$ and the first $\pi$ pulse. Approximating the $\sigma_z^{(i)}$ time dependence as a square wave, considering only the fundamental frequency (the other frequencies are at harmonics of $\omega_r$ and are therefore strongly detuned from the resonator), and in the frame rotating at the resonant spin drives the Hamiltonian is 
\begin{equation}
    \mathcal{H}/\hbar = \frac{4}{\pi}\sin{(\omega_r t + \phi)}\lambda S_z \otimes (a + a^{\dagger}) + \omega_r a^{\dagger}a.
\end{equation}

Going to the frame $U = e^{-i\omega_r a^\dagger a t}$, and setting $\phi = \pi/2$ corresponding to standard XY8-k timing, we arrive at an effective Hamiltonian in the interaction picture of: 
\begin{equation}\label{heff}
\mathcal{H}/\hbar = \frac{2}{\pi} \lambda \left(\sigma_z^{(1)} + \sigma_z^{(2)}\right) \otimes (a^{\dagger}+a),
\end{equation}
where here the pre-factor of $\frac{2}{\pi}$ comes from the Fourier transform of a square wave. Additionally, we neglect counter propagating terms rotating at frequency $2\omega_r$. Equation \eqref{heff} gives the effective Hamiltonian during the total interaction time $\tint$, i.e.\ the length of the entire pulse sequence.

\section{Semi-classical equations of motion} 
As the entanglement protocol can tolerate a highly excited thermal state of the mechanical oscillator, it is instructive to develop semi-classical equations of motion, where the bosonic mode is treated classically. Using a classical description of position and momentum $x \equiv z_p \langle a + a^\dagger \rangle$ and $p \equiv (z_p / m\omega_r) i \langle a^\dagger - a \rangle$ ($z_p$: ground state fluctuations, $m$: effective mass of the oscillator), the system undergoes the following equations of motion (without any measurements): 
\begin{equation}\label{x_evol}
\frac{dx(t)}{dt} = p(t)/m
\end{equation}
\begin{equation}
\frac{dp(t)}{dt} = -m \omega_r ^2 x(t) - \kappa p(t)  - \frac{4S_z}{\pi} \frac{\hbar \lambda}{z_p} \sin{(\omega_r t + \phi)}\Theta(t)\Theta(t-\tint) + \xi(t)
\end{equation}
Here we have assumed that the timing $2\tau$ between the $\pi$ pulses is equal to $\pi/\omega_r$, $\lambda$ is the spin-mechanical coupling strength, and we assume that the spins are in an eigenstate, such that $S_z$ is either  $0$ or $\pm 1$. We have additionally only retained the fundamental frequency of the square wave for the spin drive, since it will be the only term that provides a significant effect. Lastly, $\xi(t)$ is a Gaussian random variable that has the statistical properties 
\begin{equation}
\langle \xi(t) \rangle = 0
\end{equation}
\begin{equation}
\langle \xi(t) \xi(t') \rangle = 2D \delta(t-t'),
\end{equation}
where $D = \kappa k_B T m$ is the thermal diffusion constant ($k_B$: Boltzmann constant, $T$: temperature $\gg \hbar\omega_r/k_B$). In a viscous damping approximation and in the rotating frame (using a tilde to denote rotating frame) we find
\begin{equation}\label{x_p_rot}
\begin{split}
    \frac{d\tilde{x}}{dt} = -\frac{\kappa}{2} \tilde{x} - 2\cos{(\phi)} z_p \frac{2\lambda S_z}{\pi}\Theta(t)\Theta(t_{int}-t) + \frac{z_p^2 \bar{f}_x(t)}{\hbar} \\
    \frac{d\tilde{p}}{dt} = -\frac{\kappa}{2} \tilde{p} -\sin{(\phi)} \frac{2\hbar \lambda S_z}{z_p\pi}\Theta(t)\Theta(t_{int}-t) +  \frac{\bar{f}_p(t)}{2},
    \end{split}
\end{equation}
where we use the noise terms 
\begin{equation}
\begin{split}
    \bar{f}_x \equiv \bar{\xi}(t) + \bar{\xi}^\dagger (t) \\
    \bar{f}_p \equiv -i(\bar{\xi}(t) - \bar{\xi}^\dagger (t)),
\end{split}
\end{equation}
defined by the coarse-grained average $\bar{\xi}(t)$ of $\xi(t)$ over a short time $1/\omega_r \ll \delta t \ll 1/\kappa$. Formally integrating, the solution to equations \eqref{x_p_rot} (for $\phi = \pi/2$) is: 
\begin{equation}\label{formal_int_x_p}
\begin{split}
    \tilde{p}(t)
    = \tilde{p}(0)e^{-\kappa t/2} - \frac{4\hbar\lambda S_z}{\kappa \pi z_p}(1-e^{-\kappa t/2}) +  \int_{0}^t \frac{\bar{f}_p(t')}{2}e^{-\kappa (t-t')/2} dt' \\
    \tilde{x}(t) = \tilde{x}(0)e^{-\kappa t/2} + \frac{z_p^2}{\hbar}\int_{0}^t \frac{\bar{f}_x(t')}{2}e^{-\kappa (t-t')/2} dt'.
\end{split}
\end{equation}
The second term in the momentum equation includes the spin-dependent force responsible for our entanglement mechanism. 

Intuitively, the trajectories for different $S_z$ values must separate faster than the random walk from the diffusion to allow for a mechanical measurement of the spin state. For short times $t\ll 1/\kappa$, $\sqrt{\Delta \tilde{p}(t)^2}$ increases as $\propto\sqrt{t}$, as opposed to the expectation value $\langle\tilde{p}\rangle\propto t$ which is displaced linearly in phase space with time $t$ as shown above, such that the mechanical spin readout improves with increasing integration times, as expected.

\section{Quantum mechanical picture: effects from the environment}

In order to consider the finite thermal occupation of the resonator and the phonon induced spin decoherence associated with it, in this section we confirm that there is no phonon induced spin decoherence in the decoherence free subspace (DFS) by integrating the master equation to first order in time. For reference on the notation, see Appendix L of \cite{schuetz_high-fidelity_2017}.

In a realistic system, the resonator will be coupled to a thermal bath and the qubits will be subject to decoherence. We neglect spin flip errors (see below), and assume that the total system can be described by a master equation
\begin{equation}
    \dot{\rho}(t) = -i[\mathcal{H}, \rho] + \kappa (n_{th} + 1)\mathcal{D}(a)\rho + \kappa n_{th}\mathcal{D}(a^{\dagger})\rho + \frac{\Gamma}{2}\sum_{i = 1, 2}\mathcal{D}(\sigma_i^z)\rho.
\end{equation}
The Hamiltonian $\mathcal{H}$ is given in equation \eqref{heff} and the superoperator $\mathcal{D}$ of a generic operator $\mathcal{O}$ is $\mathcal{D}(\mathcal{O}) = \mathcal{O} \rho \mathcal{O}^\dagger - \frac{1}{2}\left(\mathcal{O}^\dagger \mathcal{O} \rho +  \rho \mathcal{O}^\dagger \mathcal{O}\right)$,  describing the rethermalization to a bath with temperature $T$ at rate $\kappa = \omega_r / Q$, such that the thermal occupation is $n_{th} = (e^{\omega_r / k_B T}-1)^{-1} \gg 1$. 
The quality factor $Q$ of the resonator is assumed to be much greater than $1$, which is well satisfied in various systems, reaching $10^9$ in some systems \cite{tsaturyan_ultracoherent_2017, ghadimi_elastic_2018}. 
We neglect depolarization of the spins and only include dephasing at rate $\Gamma = 1/T_2^*$ where $T_2^*$ is the spin coherence time: for NV centers, the depolarization timescale $T_1$ is much longer than the dynamics we consider in this proposal. Although the optimal time and fidelity may be adjusted for non-Markovian spin reservoirs such as a $^{13}C$ nuclear bath \cite{abobeih_one-second_2018}, for simplicity, we continue with a Markovian reservoir model such that the spin coherence decays exponentially. This approximation is conservative with respect to the achievable entanglement fidelities, and enables us to treat the system analytically.

\subsection{Master equation including resonator rethermalization only}

To determine on the effect of the rethermalization of the resonator, we begin with the master equation
\begin{equation} \label{master}
\dot{\rho} = -i[\mathcal{H}, \rho] + \kappa(n_{\text{th}} + 1)\mathcal{D}(a)\rho + \kappa n_{\text{th}}\mathcal{D}(a^{\dagger})\rho,
\end{equation}
where $\kappa = \omega_r/Q$ is the energy decay rate of the resonator. The solution to the master equation can be retrieved via formal integration
\begin{equation}\label{mastereqint}
\tilde{\rho}(t) = \tilde{\rho}(0) + \sum_j \gamma_j \int_0^{t} d\tau \mathcal{D}\left(\tilde{L}_j(\tau)\right) \tilde{\rho}(\tau),
\end{equation}
where here the sum is over all jump operators $\tilde{L}_j=\left\{\tilde{a}, \tilde{a}^\dagger \right\}$,   and the tilde denotes operator or density matrix in the frame $U = e^{-i \mathcal{H}t}$. 

It is instructive to examine the master equation in a rotating and displaced frame. Next, we determine how the jump operators $a$ and $a^{\dagger}$ evolve in the interaction picture, during the pulse sequence. There are two time dependent unitary transformations: To arrive at the effective Hamiltonian \eqref{heff}, we first translated into the frame given by unitary $U_1 = e^{-i\omega_r t a^{\dagger}a}$. This rotation sends the jump operators to
\begin{equation}
    a \rightarrow \tilde{a} = a e^{-i\omega_r t};\quad\quad
    a^{\dagger} \rightarrow \tilde{a}^{\dagger} = a^{\dagger} e^{i\omega_r t}.
\end{equation}
Then the spin is modulated with $\pi$ pulses, and a rotating wave approximation is made. We are left with the effective Hamiltonian \eqref{heff}, which now must \textit{also} be eliminated using the secondary unitary transformation $U_2 = e^{-i H_{\text{eff}}t}$. Since this unitary is a displacement operator, the jump operators now evolve as 
\begin{equation} \label{evolvejump}
    a \rightarrow \tilde{a} \equiv \left(a -i \frac{2}{\pi}\lambda t S_z\right)e^{-i\omega_r t}.
\end{equation}
This is a momentum displacement of $\sim \sqrt{2}\frac{2}{\pi}\lambda t S_z$, in the rotating frame, such that $\tilde{a}$ is approximately invariant under time evolution.

\subsection{Integrating the master equation}

We formally integrate \eqref{mastereqint}, to first order in time, implementing the evolved jump operators given in \eqref{evolvejump}, and noting that $\tilde{\rho}(0) = \rho(0)$ to calculate
\begin{equation}
    \tilde{\rho}^{(1)}(t) = \kappa \int_0^{t} d\tau \left(\left(n_{\text{th}} + 1\right)\mathcal{D}\left( \left(a - i\frac{2}{\pi}\lambda \tau S_z\right)e^{-i\omega_r \tau}\right) + n_{\text{th}} \mathcal{D} \left(\left(a^{\dagger} + i\frac{2}{\pi}\lambda S_z \tau\right) e^{i\omega_r \tau}\right) \right) \tilde{\rho}^{(0)}(\tau).
\end{equation}
After performing the integral, we find three types of terms, such that  such that $\rho^{(1)}(t)/t \sim {\bf A + B + C}$. First, there is a Lindbladian depicting the typical rethermalization of the resonator
\begin{equation}
    {\bf A} \propto \kappa(n_{\text{th}}+1)\mathcal{D}(a)\rho^{(0)} + \kappa n_{\text{th}}\mathcal{D}(a^{\dagger})\rho^{(0)}.
\end{equation}
This noise is to be expected, as the spin-resonator interaction does not mitigate resonator rethermalization noise. Next, we find phonon-induced decoherence of the spin
\begin{equation}
{\bf B} \propto \kappa(n_{\text{th}}+1)\mathcal{D}\left(-i \frac{2}{\pi}\lambda t S^z\right) \rho^{(0)} + \kappa n_{\text{th}}\mathcal{D}\left(i\frac{2}{\pi}\lambda t S^z\right)\rho^{(0)},
\end{equation}
such that the spins accumulate a random phase of order $\sim\lambda t$ as the resonator experiences re-thermalization noise. For the subspace corresponding to $\langle S_z \rangle =0$, this term is zero, as expected for our DFS. Lastly, there are terms that generate off-diagonal matrix elements of the resonator density matrix,
\begin{equation}
    {\bf C} \propto \kappa n_{\text{th}}\lambda t \Big(a S^z \rho^{(0)}, a^{\dagger} S^z \rho^{(0)}, S^z \rho^{(0)}a^{\dagger}, S^z \rho^{(0)}a \Big).
\end{equation}
Through the spin-resonator interaction, the resonator coherence of order $\lambda t S_z$ is injected by the spins. This coherence then rethermalizes with single phonon re-thermalization rate $\kappa n_{\text{th}}$. Note that again in the DFS, these terms are eliminated since $S_z$ is zero. 

For all three types of noise, if we include terms higher order in time, each will also be proportional either to the identity operator on the spins, $S_z$, or (for $S> 1/2$) higher powers of $S_z$, which do not cause decoherence in the DFS. This is expected: we have neglected the spins' inherent $T_1$ processes in our analysis, and both the Hamiltonian and jump opereators are proportional to $S_z$, but not $S_x$ or $S_y$. Note also that for spin states that have the same $S_z$ value but $S_z \neq 0$ (e.g., using NV $m_s = 0$ and $m_s = +1$ states as a qubit), the phonon induced phase will result in a random global phase on the spins state after projecting into $\ket{01}, \ket{10}$ subspace, such that $S_z \neq 0$ states can in principle also be employed.

\section{Entanglement heralding rates and success probability}

In the sections below, we define and discuss the following quantities in detail: 

\begin{itemize}
    \item The false positive heralding rate $r_f$: for an arbitrary attempt, the false positive $\langle S_z \rangle =0$ assignment probability (\emph{not} conditional on acceptance of the mechanical measurement as a heralding event of the spin entanglement).
    \item The true positive entanglement heralding rate $r_p$: for an arbitrary attempt, the true positive $\langle S_z \rangle = 0$ assignment probability (\emph{not} conditional on acceptance), as  defined in the main text.
    \item Probability of success $S$: \emph{given} acceptance, the probability of a true positive assignment (i.e., $\langle S_z \rangle = 0$). In the limit that the spin decoherence rate $\Gamma \rightarrow 0$, this is equal to the entanglement fidelity.
\end{itemize}

\subsection{False positive heralding rate \texorpdfstring{$r_f$}{rf}}

The false positive rate is the area under the probability density function $\mathcal{P}_{\langle S_z\rangle}(\Delta M)$ (as defined in the main text) within the thresholds and for $\langle S_z \rangle \neq 0$, weighted by the spin populations, such that 
\begin{equation}\label{tot_error}
    r_f(t) = \frac{1}{4}\text{Erfc}\left(\frac{(2-\alpha)g(t)}{\sqrt{2}}\right) - \frac{1}{4}\text{Erfc}\left(\frac{(2+\alpha)g(t)}{\sqrt{2}}\right),
\end{equation}
with $g(t)$ given in the main text, and where $\text{Erfc}(\cdot)$ is the complement of the error function. A few remarks about this result:
\begin{itemize}
    \item Note that $r_f$ is small for small values of $g(t)$: since the displacement is very small, the threshold bounds are very narrow and we almost never choose to post-select and almost never create entangled states. 
    \item Indeed, $r_f$ starts small at low $t$, then peaks as the post-selection rate goes up, and decreases again as the distributions separate. 
    \item As noted in \cite{chiani_new_2003}, the error function complement is bounded by $\text{Erfc}(x) \leq e^{-x^2}$, such that we arrive at a bound on the error
\begin{equation}\label{error_bound}
    r_f(t) \leq \frac{1}{4}e^{-(2-\alpha)^2 g(t)^2/2} \approx \frac{1}{4}e^{-(2-\alpha)^2 4C \Gamma t/\pi^2},
\end{equation}
where the last equality is for $\kappa \tint \ll$ and $\Delta m \ll \sqrt{\kappa n_{th} \tint}$ as in the main text. The false positive heralding rate is thus at worst exponentially suppressed by $C$ for fixed $t$.
\end{itemize}

\subsection{True positive entanglement heralding \texorpdfstring{$r_p$}{rp} }

This rate is given by the integrals of the probability density function $\mathcal{P}_{\langle S_z\rangle}(\Delta M)$ (as defined in the main text) within the thresholds and for $\langle S_z \rangle = 0$, multiplied by the probability of $\langle S_z \rangle = 0$ (which is equal to 1/2, given by the populations of the spin states), 
\begin{equation}
    r_p = \frac{1}{2}\text{Erf}\left(\frac{\alpha g(t)}{\sqrt{2}}\right)\geq \frac{1}{2}\left(1-e^{-\alpha^2 g(t)^2/2}\right).
\end{equation}
The bound on the right hand side of the equation is obtained from the bound on the error function complement used above. 

\subsection{Probability of success \it{S}}

\begin{figure}
\centering
\includegraphics[width=0.5\textwidth]{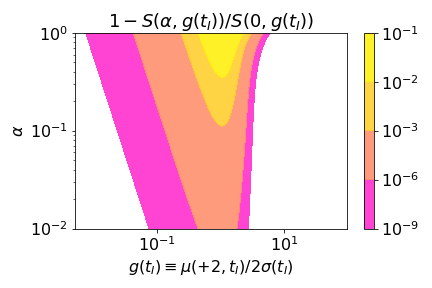}
\caption{The success probability $S(\alpha, g(t_I))$ compared to $S(0, g(t_I)$ as given in the main text.  White space corresponds to values less than $10^{-9}$. Maximum value in this plot is about 0.08.}
\label{fig:approx_S}
\end{figure}

The probability of success $S$ is given by the true positive entanglement heralding rate, normalized to the total rate of acceptance
\begin{equation}\label{fidelity}
    S = r_p/r = r_p/(r_p + r_f).
\end{equation}
We expand this expression in the threshold parameter $\alpha$ and find for small $\alpha\ll1$
\begin{equation}\label{approx_S}
    S(\alpha) \approx \frac{1}{1+e^{-2g(t)^2}} - \frac{2e^{2g(t)^2}g(t)^4}{3(1+e^{2g(t)^2})^2}(\alpha)^2.
\end{equation}
The first order term vanishes
as expected by symmetry, and the negative second order term indicates that $\alpha\rightarrow 0$ maximizes the fidelity of the entangled state. Note that $\partial^2 S/\partial \alpha^2\ll 1$ for all $g(t)$ and
$S(\alpha) \approx \frac{1}{1+e^{-2g(t)^2}}$ is an excellent approximation for many decades of $\alpha$ and $g(t)$, c.f\ Fig \ref{fig:approx_S}.
\subsection{Approximation of the optimal interaction time} 
We can describe the optimal interaction time $\tint$ by optimizing the fidelity given in the main text. For $\alpha \rightarrow 0$
\begin{equation}
    \frac{d \mathcal{F}}{d (\Gamma t)} = \frac{8C e^{-16C \Gamma t / \pi^2}(1+e^{-2 \Gamma t})}{\pi^2 (1+e^{-16 C\Gamma t/\pi^2})^2} - \frac{e^{-2\Gamma t}}{1+e^{-16 C \Gamma t/\pi^2}}.
\end{equation}
To approximate the optimal time $\topt$, we assume that $\Gamma \topt \ll 1$ and expand $e^{-2\Gamma t}$ to zeroth order, setting $e^{-2\Gamma t} \approx 1$, and do not expand the exponential $e^{-16 C \Gamma t/\pi^2}$ since $C$ can be large. We arrive at the expression
\begin{equation}
    \Gamma \topt \approx \frac{\pi^2 \text{ln}(16C/\pi^2 -1)}{16C},
\end{equation}
as stated in the main text. We note that we assumed a Markovian spin bath such that the spin coherences decay as a simple exponential. A non-Markovian spin reservoir (as encountered in dense spin ensembles, e.g., for NV centers coupled to a bath of $^{13}$C nuclear spins), with same $\Gamma$ would result in better performance, since $\Gamma \topt \ll 1$, but cannot be approximated analytically in the same manner.

\section{Inhomogeneous coupling strength}
We can include a small difference in coupling $\Delta \lambda$ by treating the DFS $\mathscr{H}_{DFS} \equiv \left\{\frac{\ket{01}+\ket{10}}{\sqrt{2}}, \frac{\ket{01}-\ket{10}}{\sqrt{2}}\right\}$ as an effective two-level system coupled to a resonator, undergoing thermal noise, with coupling strength $\Delta \lambda$, such that the effective Hamiltonian is
\begin{equation}
    \mathcal{H}/\hbar = \frac{2}{\pi}\lambda S_z \left(a + a^{\dagger}\right) + \frac{2}{\pi}\frac{\Delta \lambda}{2} \left(\sigma_z^{(1)}-\sigma_z^{(2)}\right)\left(a + a^{\dagger}\right).
\end{equation}
We describe three contributions to the reduction in fidelity from a finite $\Delta \lambda$: (i) phonon induced spin decoherence, a random phase accumulated onto the two spin coherences given by the random fluctuations of the resonator state, as the spin senses the resonator position, 
(ii) the information gained by the measurements on the resonator that projects the spins onto the $\ket{01}$ or $\ket{10}$ state when $\langle S_z \rangle = 0$, and
(iii) the difference in the spin-induced momentum displacements from the ideal case $\Delta \lambda = 0$, given by the four possible forces exerted on the resonator, 
In the following, we treat each of these three effects independently, assuming that the displacement $\sim \Delta \lambda \tint$ is much smaller than the diffusion $\sqrt{\kappa n_{th} \tint}$. 

\subsection{Phonon induced spin decoherence} 
\begin{figure}
\centering
\includegraphics[width=0.7\textwidth]{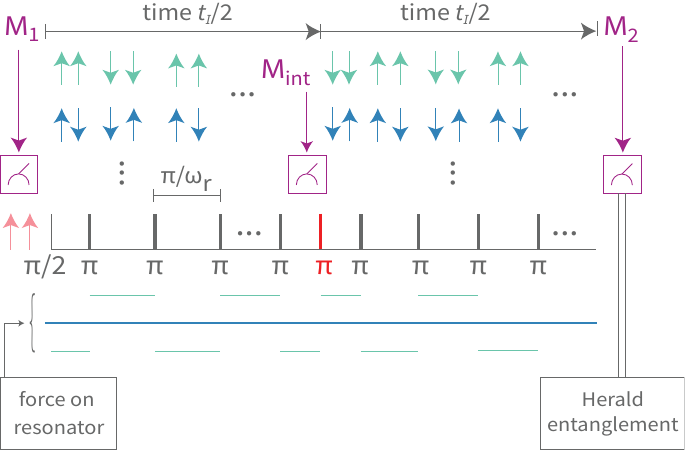}
\caption{Pulse sequence replacing that of figure 1(c) that mitigates phonon-induced decoherence in the presence of inhomogeneous coupling. The extra $\pi$ pulse (red) halfway through the interaction time $\tint$ eliminates much of the decoherence due to the resonator phase.}
\label{fig:x_gate}
\end{figure}

Without additional spin control to mitigate the phonon-induced spin decoherence, the spin $\langle S_z \rangle = 0$ state will become an incoherent mixture of $\ket{\Psi^+}$ and $\ket{\Psi^-}$, reducing the spin coherence according to the decrease in spin contrast shown for a thermal resonator state in \cite{kolkowitz_coherent_2012}, which can drastically reduce the fidelity. However, since the duration of our sequence $t \ll 1/\kappa$, then the signal is largely coherent during the pulse sequence, such that we can apply dynamical decoupling within $\mathscr{H}_{DFS}$ to suppress its effect. If we add an intermediate measurement $M_{int}$ at time $\tint/2$,  threshold on the variable $\Delta M_{new} \equiv M_2 - M_{int}e^{-\kappa \tint/4} - (M_{int}e^{-\kappa \tint/4} - M_1e^{-\kappa \tint/2})$, and add an additional $\pi$ pulse at time $\tint/2$, we dynamically decouple the coherent part of the resonator signal (see figure \ref{fig:x_gate} for an illustration). This is equivalent to applying a so-called Hahn echo sequence in $\mathscr{H}_{DFS}$, while retaining the same optimal time $\topt$. In our analysis below, for simplicity we again assume that the measurement uncertainty is negligible. 

In this section, we quantify the remaining random phase and decoherence after such a sequence due to the phonon-induced spin thermalization, as well as a new success probability defined by all three measurements. As mentioned above, we assume $\Delta \lambda \tint \ll \sqrt{\kappa n_{th} \tint}$, such that we can use a semi-classical picture, neglecting the small (coherent) phase added to the spins from the difference in the spin-induced resonator displacements between the $\ket{01}$ and $\ket{10}$ states. 
A simple estimate first provides a quite accurate result. During the evolution over the pulse sequence of time $\tint$, diffusion occurs with variance given by the single phonon decay rate, $\kappa n_{th} \tint$. As this evolution is purely Markovian, the extra $\pi$ pulse halfway through the sequence does not prevent the spins from sensing these fluctuations with coupling strength $\Delta \lambda$. Thus we expect the two qubit coherences to accumulate a random phase $\sim \Delta \lambda \tint \sqrt{\kappa n_{th} \tint}$.

More precisely, we describe a two-level system $\ket{\psi}_{DFS}(t)$ in $\mathscr{H}_{DFS}$ in the rotating and toggling frame (eq. \eqref{heff}) that begins in the state $\ket{\psi(t)}_{DFS} = \ket{\Psi^+}$ and accumulates phase during the pulse sequence over time $\tint$ such that, up to a global phase the state becomes
\begin{equation}\label{psi_dfs_first_half}
\ket{\psi(\tint/2)}_{DFS} = \frac{\ket{10} + e^{i\phi(\tint/2)}\ket{01}}{\sqrt{2}},
\end{equation}
where the oscillator-state-dependent phase $\phi(t)$ is given by
\begin{equation}
    \phi(t) = \frac{4}{\pi}\frac{\Delta \lambda}{z_p} \int_0^t \tilde{x}(t') dt',
\end{equation}
and $\tilde{x}(t)$ is described by equation \eqref{formal_int_x_p}. Note that to simplify the derivation, we have again made the same rotating wave approximation and neglected the higher harmonics of the square wave, unlike the result in \cite{kolkowitz_coherent_2012} (in the limit $\kappa \tint \ll 1$, the results are equivalent). After the extra $\pi$ pulse on both spins, and further evolution for another time $\tint/2$, the state is
\begin{equation}\label{psi_dfs_second_half}
\ket{\psi}_{DFS}(t) = \frac{\ket{10} + e^{i\Delta \phi(\tint)}\ket{01}}{\sqrt{2}},
\end{equation}
where
\begin{equation}
    \Delta \phi(t) = \frac{4}{\pi} \frac{\Delta \lambda}{z_p} \Big(\int_{t/2}^t x(t') dt' - \int_{0}^{t/2} x(t') dt'\Big).
\end{equation}
Taking the statistical average over the distribution for $\Delta \phi(t)$, we calculate the spin coherences of the density matrix to be
\begin{equation}
    \langle e^{\pm i\Delta \phi(t)} \rangle = e^{-\sigma_{\Delta \phi}^2(t)/2},
\end{equation}
where $\sigma^2_{\Delta \phi}(t)$ is the variance of $\Delta \phi(t)$, as $\tilde{x}(t)$ is Gaussian. Thus, neglecting the change in the distributions of the resonator state due to the inhomogeneous coupling (treated independently below), the fidelity is reduced by a factor (see also the next section)
\begin{equation}
    \mathcal{F} \rightarrow \mathcal{F}\cdot\left(\frac{1 + e^{-\sigma_{\Delta \phi}^2(\tint)/2}}{2}\right).
\end{equation}
To calculate $\sigma_{\Delta \phi}^2(t)$, we apply equation \eqref{formal_int_x_p} and integrate:
\begin{align}\nonumber
    \sigma_{\Delta \phi}^2(t) &= \left(\frac{4}{\pi} \frac{\Delta \lambda z_p}{\hbar} \int_{t/2}^t \int_{0}^{t'} \frac{\bar{f}_x(\tau_1)}{2}e^{-\kappa (t'-\tau_1)/2} d\tau_1 dt'
    - \frac{4}{\pi} \frac{\Delta \lambda z_p}{\hbar} \int_{0}^{t/2} \int_{0}^{t''} \frac{\bar{f}_x(\tau_2)}{2}e^{-\kappa (t''-\tau_2)/2} d\tau_2 dt''\right)^2 \\
    &= \frac{16}{\pi^2}\frac{\Delta \lambda^2 z_p^2}{\hbar^2}\frac{8D}{\kappa^3}\left(\kappa t + 12 e^{-\kappa t/4} + 4e^{-3\kappa t/4}- 8 e^{-\kappa t/2} - e^{-\kappa t} -7\right),
\end{align}
where in the second line, $D$ is the diffusion constant. If we take the limit that $\kappa t \ll 1$ and expand in small $\kappa t$ to the lowest non-zero order (third order in $\kappa t$) we find that
\begin{equation}\label{phonon_induced_decoh}
\sigma^2_{\Delta \phi}(t) \approx \frac{16}{3\pi^2} \kappa n_{th} \Delta \lambda^2 t^3
\end{equation}
such that the average phase induced $\sqrt{\sigma^2_{\Delta \phi}(t)} \sim \Delta \lambda t \sqrt{\kappa t n_{th}}$ as predicted above. We can calculate a maximum allowed $\Delta \lambda_{max, \phi}$ such that this phonon-induced decoherence does not limit the fidelity for a given error rate $\mathcal{E}$ (see fig. 3, equation 8 of the main text), by requiring that the error contributed by the phonon induced decoherence is at most the size of the original error. The maximum allowed inhomogeneity is then defined to be
\begin{equation}
    \Delta \lambda_{max, \phi} \sim \Big(\frac{3\pi^2}{16 \kappa n_{th} \tint^3}\text{ln}\frac{1}{(1-2\mathcal{E}_T)^2}\Big)^{1/2}.
\end{equation}
Note that additional measurements and subsequent feedback (through additional electron spin control or restarting if $S_z \neq 0$ is likely) can augment this procedure to increase $\Delta \lambda_{max, \phi}$, beyond the scope of this work.

\subsection{Projection into \texorpdfstring{$\{\ket{10}, \ket{01}\}$}{|10>, |01>} states}
When we perform the pulse sequence for time $\tint$ and measurements in the presence of an inhomogeneity in the coupling strength, there is some information gain about whether $\ket{10}$ or $\ket{10}$ is populated, reducing the entanglement fidelity. 

Here, we show an optimal reconstruction of spin state after a particular measurement $z$, calculate the fidelity of the new state with the $\ket{\Psi^+}$ state, and then average over all measurements $z$, weighted by its Gaussian distributions. We then require that the error caused by this effect is less than the entanglement preparation error $\mathcal{E}$, to define a new $\Delta \lambda_{max, m}$ given by the measurement and subsequent projection. 
As we will see in the following, this process critically depends on a finite measurement uncertainty $\Delta m\neq 0$. Note that this is not in contradiction with the previous assumption $\Delta m \ll \kappa n_{th} \tint$. 

We define state $\ket{a} \equiv \ket{01}$ and $\ket{b} \equiv \ket{10}$, and calculate our result using the effective two level system $\mathscr{H}_{DFS} \equiv \{\ket{a}, \ket{b}\}$.

Suppose we measure the displacement and obtain value $z$. The optimal estimates for the populations of the spin states are 
\begin{equation}
    \rho_{a(b)} = \frac{\mathcal{P}_{a(b)}}{\mathcal{P}_{a} + \mathcal{P}_{b}},
\end{equation}
where $\mathcal{P}_{a (b)}(z)$ is the probability distribution for measurement $z$ given state $a(b)$, a normal distribution with mean $\pm \delta g$ and variance $\sigma^2(\tint)$ as defined in the main text. Since the measurement only yields information about the population of the spin eigenstates and not the coherences, we assume that our measurement is projective only along the z axis of $\mathscr{H}_{DFS}$. Thus, after the measurement with value $z$, the density matrix will be
\begin{equation}
    \rho(\tint | z) = \begin{pmatrix} \rho_a(z) & \rho_{ab}(z) \\ \rho_{ab}(z) & \rho_b(z) \end{pmatrix},
\end{equation}
where we approximate the coherence as $\rho_{ab}(z) = e^{-\sigma^2_{\Delta \phi}(t)/2}\sqrt{\rho_{a}(z)\rho_{b}(z)}$, the maximum, real value, since the state in $\mathscr{H}_{DFS}$ begins in $\ket{+\text{x}}$ and the measurement only weakly projects along the $z$ axis. 
The prefactor of the coherence (see previous section) can be interpreted as the information on the spin state learned by the environment during the pulse sequence. To fully separate these two processes (information gain by the environment and by the measurement), we formally assume the back-action of the environment on the mechanical resonator (i.e.\ it's diffusion) is known an can be subtracted from the physical measurement result, such that uncertainty in the difference measurement $\Delta \bar{m}$ is only given by the uncertainty of the resonator measurements $\Delta m$. For example, the sequence of figure \ref{fig:x_gate} results in $\Delta \bar{m}^2 \approx 6 \Delta m^2$ for three identical measurements. 
Integrating equation \eqref{formal_int_x_p}, we obtain the two $\langle S_z \rangle =0$ displacements $\delta \mu \approx \pm 2\sqrt{2}\Delta \lambda t/\pi$ for $\Delta \bar{m}^2\ll\sigma^2(\tint) \approx \kappa n_{th} \tint$.
The fidelity then becomes
\begin{align}
\mathcal{F} &\rightarrow \mathcal{F} \cdot \left( \frac{1}{2} + \frac 1 2 e^{-\sigma^2_{\Delta \phi}(t)/2}\int_{-\infty}^{\infty} \sqrt{\mathcal{P}_a(z)\mathcal{P}_b(z)} \text{ d}z \right) \\
&= \mathcal{F} \cdot \left(\frac{1}{2} + \frac{e^{-\sigma^2_{\Delta \phi}(t)/2}}{2} \int_{-\infty}^{\infty}\frac{\sqrt{e^{-(z-\delta \mu)^2/2\Delta \bar{m}^2}e^{-(z+\delta \mu)^2/2\Delta \bar{m}^2}}}{\sqrt{2\pi}\Delta \bar{m}} \text{ d}z\right) \\
&\approx \mathcal{F} \cdot \left(1-\frac{\delta \mu^2}{4\Delta \bar{m}^2} - \frac{\sigma^2_{\Delta \phi}(t)}{4}\right),
\end{align}
where in the last line we have expanded to lowest non-zero order in $\delta \mu$ and $\sigma^2_{\Delta \phi}(t)$. The second (third) term describes the infidelity associated with the measurement (phonon induced decoherence, treated above).
To ensure $\frac{\delta \mu^2}{4\Delta \bar{m}^2}\ll \mathcal{E}_T$, we find
the maximum inhomogeneity $\Delta \lambda_{max, m}$ which does not limit the fidelity is
\begin{equation}
    \Delta \lambda_{max, m} \sim \left( \frac{\pi^2 \Delta \bar{m}^2 \mathcal{E}}{2\tint^2}\right)^{1/2}.
\end{equation}

\subsection{Change in resonator displacement due to \texorpdfstring{$\Delta \lambda$}{Delta lambda}}

Here, we estimate the effect of the slight displacement of the resonator $\delta \mu$  on the fidelity through the reduction in the success rate $S$, and calculate the maximum allowed $\Delta \lambda_{max, disp}$ such that this effect does not limit the error. With the differential normalized displacement $\delta g \equiv \delta \mu / \sigma$,
the true positive entanglement heralding rate $r_p$ as defined in the main text is
\begin{equation}\label{dlambda_s}
r_p(\delta g) = \frac{1}{2} - \frac{1}{4}\left( \text{Erfc}\left(\frac{\alpha g + \delta g}{\sqrt{2}}\right) + \text{Erfc}\left(\frac{\alpha g - \delta g}{\sqrt{2}}\right)\right),
\end{equation}
with $\sigma \approx \kappa n_{th} t$ as defined in the main text, and the false positive heralding rate $r_f$ is unchanged. For a minimal change in the error $\mathcal{E}$ we require that the change in success rate $|\delta S(\alpha, g, \delta g)/| \ll \mathcal{E}$, yielding the condition, by expanding to highest order in $\delta g$ that
\begin{equation}
    \frac{\delta g^2 \alpha g e^{-\alpha^2 g^2 /2}}{2\sqrt{2\pi} \left(r_{p0} + r_f \right)^2}\ll \mathcal{E},
\end{equation}
where $r_{p0}$ is the true positive rate when $\delta g = \Delta \lambda = 0$. This leads to a condition on the maximum allowed coupling strength $\Delta \lambda_{max, disp}$ of
\begin{equation}
    \Delta \lambda_{max, disp} \sim \frac{\pi\sigma (r_{p0} +r_f)}{2\sqrt{2}\tint} \left(\frac{2\sqrt{2\pi} \mathcal{E} e^{\alpha^2 g^2 /2}}{\alpha g}\right)^{1/2}.
\end{equation}
\subsection{Summary of limits on inhomogenous coupling}
We define the maximum allowed inhomogeneity that does not limit the error to be
\begin{equation}
    \Delta \lambda_{max} = \text{min}(\Delta \lambda_{max, \phi}, \Delta \lambda_{max, disp}, \Delta \lambda_{max, m}).
\end{equation} 
We find that for our parameter regions of interest and in table \ref{tab:table}, $\Delta \lambda_{max, \phi} < \Delta \lambda_{max, disp}, \Delta \lambda_{max, m}$, as expected from the assumption that the system is diffusion limite. The maximum inhomogeneity is on the order of a percent of $\lambda$ (see table \ref{tab:table} for a list of $\Delta \lambda_{max}$ associated with each example parameter set). However, for extremely small $\alpha$, $\Delta \lambda_{max, disp}$ can be small. Note that if $\Delta \lambda$ is known, the thresholds may be reoptimized to improve $\Delta \lambda_{max, disp}$. As discussed above, the phonon induced decoherence $\sigma_{\Delta \phi}(t)$ can be further improved with additional measurements and feedback. Furthermore, we note that for a known but finite $\Delta \lambda$, $\Delta \lambda_{max, m}$ may be improved by optimizing the measurement uncertainty $\Delta m$.

\section{Example parameters}
In the table \ref{tab:table}, we delineate various parameter sets and their performance under our entanglement mechanism. 

\begin{table}[ht!]
\centering
 \begin{tabular}{||c c c c c c c c c c c||} 
 \hline
Label & $1/\Gamma$ & $Q$ & $\lambda / 2\pi$ (Hz) & $T$ (K) & $C$ & $\Delta m^2$ & $\topt$ (ms) & $\mathcal{E}$ (\%) & $r_p$  & $\frac{\Delta \lambda_{max}}{\lambda}$ (\%)\\ [0.5ex] 
 
 \hline\hline
 1 \label{row:1} & $10$ms & $10^7$ & $450$ & $4$ & $1.5$ & $24$ & $8.9$ & $49$ & $0.16$ & 1.5\\ 
 \hline
 2 \label{row:2} & $1.6$s & $10^9$ & $100$ & $4$ & $1206$ & $8$ & $17.5$ & $1.2$ & $0.29$ & 2.3 \\ [1ex] \hline \hline
 
 3 \label{row:3} & $0.6$s & $10^9$ & $1000$ & $77$ & $2350$ & $10$ & $2.6$ & $0.48$ & $0.31$ & 0.40 \\
 
 \hline\hline
 4 \label{row:4} & $10$ms & $10^9$ & 400 & 293 & 1.6 & 20 & 8.2 & 48 & 0.15 & 1.8 \\ \hline
 
 5 \label{row:5} & $10$ms & $10^9$ & $880$ & $293$ & $8$ & $27$ & $3.1$ & $28$ & $0.18$ & 1.4 \\ \hline
 6 \label{row:6} & $10$ms & $10^{10}$ & $2000$ & $293$ & $412$ & $0.06$ & $0.14$ & $1.5$ & $0.28$ & 4.9 \\ \hline
    \end{tabular} 
    \caption{Example parameters for application of our protocol, for corresponding Bell state preparation error $\mathcal{E}$ at cryogenic temperatures (rows 1 and 2) at liquid nitrogen temperature (row 3), and room temperature (rows 4-6). The rate $r_p$ is the true positive rate, i.e., the inverse of the average number of repetitions until acceptance.
    Fifth row: example parameters using values previously cited in the literature \cite{ghadimi_elastic_2018, tsaturyan_ultracoherent_2017, rossi_measurement-based_2018, bar-gill_solid-state_2013, guo_feedback_2019}. In all cases, the threshold parameter was $\alpha = 0.4$, and the optimal time $\topt$ is found including measurement uncertainty. For simplicity, the maximum allowed inhomogeneity $\Delta \lambda_{max}/\lambda$ is calculated for the corresponding error neglecting the measurement uncertainty but with the same cooperativity and spin-resonator parameters. Note that in almost all cases, cooling to the ground state by measurement is not required, i.e. $\Delta m^2 \gg 1$. For these, a continuous-wave measurement with power $\lesssim $mW (for $\omega_r / 2\pi = 10^6$, $z_p = 10^{-14}$, corresponding to a high stress SiN beam of length $\sim 100$~$\mu$m), and a Kalman filter in the steady state \cite{hofer_quantum_2017, stengel_optimal_1994, bryson_applied_1975, hassani_further_2009} can be employed. The final row assumes a backaction evading measurement \cite{vanner_pulsed_2011}.}
    \label{tab:table}
\centering
\end{table}

\section{Estimating the resonator state with Kalman filters}

Estimating the state of the resonator is a key step in the entanglement protocol presented here. As the resonator follows linear equations of motion, and is subject to a thermal noise bath and a (spin-state conditional) deterministic force, its state can be estimated with a Kalman filter. This is an algorithm from optimal control theory that allows for reconstructing the state of a linear system subject to white noise from a series of measurements with minimal uncertainty. 

Here, we employ the framework of a Kalman filter for a realistic estimates of the power required for the resonator measurements, and analytically derive the expectation value and variance of the difference between the two measurements $M_1$ and $M_2$ as mentioned in the main text. 
In an experimental implementation, a Kalman filter can further be used to address typical deviations of experimental implementations from the simple analytical model presented here, as well as for further optimization of the protocol. For example, auxiliary mechanical modes can be included in the model of the system.
Similar to entanglement protocols for trapped ions, spectator modes can be detrimental for the fidelity of the scheme presented here if they are not accounted for. Employing a Kalman filter, the state of the spectator modes can be estimated simultaneously, enabling us to extract the momentum displacement of only the mode of interest. Furthermore, the Kalman filter can be employed in combination with multiple model adaptive estimation to perform the mechanical measurements during the pulse sequence, further decreasing the time required to establish entanglement between the spin states. As the latter two cases require various unnecessary additional assumptions and cannot be treated analytically, we focus here on the case of a single mechanical mode. 

Our protocol with Kalman filters consists of the sequence consisting of the following steps: 
\begin{enumerate}
    \item Measure the resonator state (phase and amplitude) using an interferometer. Apply a Kalman filter to the results to optimally estimate the resonator state $M_1$ with minimal, steady-state uncertainty $\Delta m$.
    \item Execute the spin pulse sequence.
    \item Measure the resonator state (phase and amplitude) using an interferometer. Apply a Kalman filter to the measurement results $M_2$, to optimally estimate the state of the resonator at time $\tint$, with minimal, steady-state uncertainty $\Delta m$.
    \item If the estimated final resonator state is within a threshold about the initial state, continue the circuit (spin entanglement has been generated with fidelity $\mathcal{F}$). Otherwise, restart.  
\end{enumerate}

As illustrated in figure 2(a) of the main text, step (1) localizes the resonator state to some Gaussian state with uncertainty smaller than the thermal distribution. Step (2) introduces shifts of the localized resonator state due to the various spin populations. Step (3) records the resonator state after the interactions with the spins. Step (4) allows for post-selection on entangled states and collapses the spin state to either a separable or entangled state, with fidelity $\mathcal{F}$. If the resonator state is approximately unchanged between  steps (1) and (3), then the spins are projected into a Bell state.

Note that we assume the measurement duration is much less than the interaction time $\tint$, such that the optimal time is accurately represented in the main text. For further description, see sections 7.4.3 and 8 below. Typically, the covariance matrix of a Kalman filter comes close to the steady state in a time given by the inverse of the smallest frequency gap in the spectrum, in our case $2\pi/\omega_r$.

\subsection{Step 1 - Kalman filter localizes the initial resonator state}

Before the measurement, the resonator begins in a thermal state, $\rho(0) = \rho_{\text{th}}$ at temperature $T$ and resonator frequency $\omega_r$. Let the measurement result value at time $t$ be $z(t)$, which contains the signal, photon shot noise, and the process noise - fluctuations and drifts from the finite $Q$ factor of the resonator and its temperature. In this section, we describe the Kalman filter and its results to localize the resonator state within the thermal distribution. For more information on the filter as well as the notation employed here, see \cite{stengel_optimal_1994} and \cite{bryson_applied_1975}.

\subsubsection{Review of Kalman filters}

We denote the classical state vector as $\bf{X} =\begin{pmatrix} x \\ p \end{pmatrix}$, and the corresponding estimated state is denoted $M = \begin{pmatrix} \hat{x} \\ \hat{p} \end{pmatrix}$. After some measurements, the system (resonator, measurement, and filter) will approach a steady state, where the information gained by each measurement is equal to the information lost by the dissipation and noise. The covariance matrix $P(t) \equiv \langle ({\bf{X}}(t) -  {M}(t))({\bf{X}}(t) - {M}(t))^{\text{T}} \rangle $ at this steady state is denoted by $P_{ss}$. The measurement uncertainty $\Delta m$ will be equal on both quadratures and correspond to the diagonal elements of $P_{ss}$. The state vector undergoes an equation of motion
\begin{equation}\label{state_vec_evol}
    {\bf{\dot{X}}}(t) = F {\bf{X}}(t) + G u(t) + L w(t),
\end{equation}
where the evolution matrix $F$ describes the resonator physics relating the two quadratures, $G$ is the deterministic drive, and $L$ and $w(t)$ describe process (diffusive) noise. The measurements are described by
\begin{equation}\label{meas}
    z(t) = H {\bf{X}}(t) + n(t),
\end{equation}
where the row vector $H$ maps the state vector onto the measurement value and $n(t)$ is a Gaussian noise process corresponding to the measurement (photon shot noise), such that $\langle n(t) n(t') \rangle = R \delta(t-t')$ and $\langle n(t) \rangle = 0$. Following \cite{stengel_optimal_1994}, let us assume that there is a particular state estimate at some point in time $t$, $\bf{X'}$. The covariance matrix of the estimate $M$ at this point in time $t$ is  $\text{E}[({\bf{X}} - {\bf{X'}})({\bf{X}} - {\bf{X'}})^T] = M$. Also, let us assume that we have a set of discrete measurements $\bf{z} = \{z_1(t), z_2(t), ... \}$ that were performed \emph{after} time $t$ at which we have the estimate $\bf{X'}$, and the relationship between each $z(t)$ in $\textbf{z}$ and the state $\bf{X}$ is given by equation \eqref{meas}. The cost function (a scalar) is the so-called \emph{weighted least-squares estimate}
\begin{equation}
    J = \frac{1}{2}\Big(({\bf{X}} - {\bf{X'}})^{T} M^{-1}({\bf{X}} - {\bf{X'}}) + ({\bf{z}}-H{\bf X}) R^{-1} ({\bf{z}}-H{\bf X}) ^{\text{T}}\Big).
\end{equation}
Optimizing the filter, taking the limit of continuous measurements, and using the notation described in \cite{stengel_optimal_1994}, we obtain the evolution equation for the estimate and the covariance matrix, respectively:
\begin{align}\label{Kal_filter_estimate}
    {\bf{\dot{X}}}(t) &= F {M}(t) + G u(t) + K (t) [z(t) - H M(t)],\\
\label{Kal_filter_cov}
    \dot{P}(t) &= F P(t)+ P(t)F^{T} + L Q L^{T} - P(t) H^{T} R^{-1} H P(t).
\end{align}
Note that the drive $G, u$ is assumed to be a known, deterministic parameter, such that the covariance does not depend on $G, u$. The so-called `Kalman gain' is: 
\begin{equation}
    K(t) = P(t) H^{T} R^{-1}.
\end{equation}
For the steady state we find
\begin{equation} \label{Pss}
    0 = F P_{ss} + P_{ss} F^{T} + L Q L^{T} - P_{ss} H^{T} R^{-1} H P_{ss}.
\end{equation}
This Lyapunov equation is a system that can be solved analytically for our 2x2 matrices in particular. Below, we evaluate the steady state matrix $P_{ss}$ in the lab frame, under a rotating wave approximation (RWA). 

\subsubsection{Derivation for \texorpdfstring{$P_{ss}$, $\Delta m$}{Pss, Dm}}

We can simplify the result for $P_{ss}$ if we allow for measurements to be performed on the resonator's momentum.  We cannot physically build an experiment to do this directly, as the phase of the reflected laser beam in the interferometer depends on the resonator position only. However, we can make an rotating wave approximation, such that waiting for time $\frac{1}{4}\frac{2\pi}{\omega_r}$ after measuring the position is equivalent to measuring the momentum: the momentum and position are highly correlated in the lab frame, such that they rotate into each other much faster than the diffusion in phase space described by the finite $Q$ and temperature.

Below, we choose units of momentum such that the $x$ quadrature is scaled by $m\omega_r$ such that $x \rightarrow m\omega_r x$, with $p$ kept the same. The evolution matrix is: 
\begin{equation}
    F = \left ( {\begin{array}{cc}
   0 & \omega_r \\
   -\omega_r & -\kappa \\
  \end{array} } \right).
\end{equation}
Since we are performing the measurements without the spin pulse sequence on, the drive term $u(t) = 0$. We only have a single process noise source here, thermal driving noise, such that $w(t)$ is a scalar and specifically is equal to $\xi(t)$ as defined above, with the noise spectral matrix $Q_c$ equal to twice the scalar diffusion constant $D$:
\begin{equation}
    \langle w(t)w(t') \rangle = 2k_B T \kappa m \delta(t-t').
\end{equation}
To describe the Brownian motion process leading to dissipation, we have: 
\begin{equation}
    L =  \left ( {\begin{array}{cc}
   0 \\
   1 \\
  \end{array} } \right).
\end{equation}
The matrix $H$ converts the state vector into the equivalent measurement and is
\begin{equation}\label{H_meas}
    H = \left (0, 
    2\pi R/(m\omega_r \lambda_l) \right).
\end{equation}
The noise $n(t)$ has properties $\langle n(t) \rangle = 0$ and $\langle n(t) n(t') \rangle = R \delta(t-t')$, where $P$ is the scattered power of the laser beam and $E$ is the energy of each photon and the photon flux rate is $R = P/E$. In this case, we have for the steady state
\begin{equation}
\begin{split}
    0 = \begin{pmatrix} 0 & \omega_r \\ -\omega_r & -\kappa \end{pmatrix} \begin{pmatrix} P_{11} & P_{12} \\ P_{12} & P_{22} \end{pmatrix} + \begin{pmatrix} P_{11} & P_{12} \\ P_{12} & P_{22} \end{pmatrix} \begin{pmatrix} 0 & -\omega_r \\ \omega_r & -\kappa \end{pmatrix} +  
    \begin{pmatrix} 0 & 0 \\ 0 & 2D \end{pmatrix}\\
     - \begin{pmatrix} P_{11} & P_{12} \\ P_{12} & P_{22} \end{pmatrix} \begin{pmatrix} 0 \\ 2\pi R/(m\omega_r \lambda_l) \end{pmatrix} \begin{pmatrix} 0 & 2\pi R/(m\omega_r \lambda_l) \end{pmatrix} \begin{pmatrix} P_{11} & P_{12} \\ P_{12} & P_{22} \end{pmatrix} \frac{1}{R}.
\end{split}
\end{equation}
Solving the system of equations we find that, in units of momentum, the measurement variance is
\begin{equation}
\begin{split}
    P_{11} = P_{22} = \left(\frac{\hbar \lambda_l}{z_p^2}\right)^2 \frac{1}{16\pi^2 R}\left(-\kappa + \sqrt{\kappa^2 + \left(\frac{2 z_p}{\lambda_l}\right)^2 4\pi^2 R \kappa n_{th}}\right),
\end{split}
\end{equation}
and converting to units of number of excitations
\begin{equation}
\begin{split}
    \Delta m^2 = \left(\frac{\lambda_l}{2\pi z_p}\right)^2\frac{1}{R}\left(-\kappa + \sqrt{\kappa^2 + 4R\kappa n_{th}\left(\frac{2\pi z_p}{\lambda_l}\right)^2}\right),
\end{split}
\end{equation}
a result consistent with \cite{bryson_applied_1975}. As expected, the correlations $P_{12}, P_{21}$ are zero, and the process and measurement noise symmetrically effect both quadratures.

The quantity in the square root, $\sim n_{th}$, is generally much larger than 1 for our parameters (it is of order $\sim 10^9$). We can therefore expand in its inverse and we find to zeroth order, in units of phonon excitations, that:
    \begin{equation}\label{Delta_m}
    \Delta m^2 \approx 2\left(\frac{\lambda_l}{2\pi z_p}\right)\sqrt{\frac{\kappa n_{th}}{R}}.
    \end{equation}
The variance of both quadratures is the resolution of the interferometer in units of the zero point motion, multiplied by the square root of the net rate of information gain by each photon as decoherence and information gain compete, namely $R/\kappa n_{th}$.

\subsubsection{A note on quantum backaction noise}

 To include quantum backaction noise on the resonator from photon shot noise, we have a new steady-state equation for the covariance, because the measurement and process noise is now correlated. However, for continuous wave Kalman filtering as described above, and as quoted in table \ref{tab:table}, note that in our parameter regime of interest, $\Delta m^2 \gg 1$, such that we can safely neglect quantum backaction.
 
 Additionally, as stated in the main text, we can also consider backaction evading measurements, strong pulses that are timed to act on a particular quadrature, thus broadening the other quadrature, and carefully time the pulse spacing to induce displacement on the quadrature that does not receive the quantum backaction noise. See also section 8 for further description and the last row of table \ref{tab:table}.

\subsection{Step 2 - the spin pulse sequence}

Next, the spin pulse sequence begins, which displaces the resonator momentum according to eq. \eqref{formal_int_x_p} over time $\tint$.

\subsection{Step 3 - measurement after spin pulse sequence}

After the spin pulse sequence, another measurement is performed, again for the (short) time required to reach the steady-state covariance $P_{ss}$. Suppose the estimate at this time $\tint$ is ${M(\tint)}$ and covariance given by $P_{ss}$. At this point, we now have two estimates: ${M(0)}$ and ${M(\tint)}$. The variance of the final position estimate is: 
\begin{equation}\label{var_final_est_pos}
    \left\langle \hat{\tilde{x}}(\tint)^2 \right\rangle - \left\langle \hat{\tilde{x}}(\tint) \right\rangle^2 = \Delta m^2 + n_{th}\left(1-e^{-\kappa \tint}\right),
\end{equation}
since the variance should be given by the sum of two independent Gaussian random variables: the position $x(t)$ after diffusion as well as the measurement noise $\Delta m^2$. \footnote{$\Delta m^2$ is in the lab frame, but our diffusion description is in the rotating frame. Because the momentum and position covariances are both equal to $\Delta m^2$, they also have covariance $\Delta m^2$ in the rotating frame, so our description is self-consistent.} By the same argument, the variance of the final momentum estimate is also
\begin{equation}\label{var_final_est_mom}
    \left\langle \hat{\tilde{p}}(\tint)^2 \right\rangle - \left\langle \hat{\tilde{p}}(\tint) \right\rangle^2 = \Delta m^2 + n_{th}\left(1-e^{-\kappa \tint}\right).
\end{equation}

\subsection{Step 4 - threshold on the \texorpdfstring{$\langle S_z \rangle =0$}{\it{Sz=0}} state}

Next, we use the two estimates $M_1$ and $M_2$ to threshold appropriately on the $\langle S_z \rangle =0$ state. To do so, we consider the difference of the two estimates, accounting for dissipation, one from $t = 0$ and one at time $\tint$: $\Delta M \equiv M_2 - e^{-\kappa \tint/2}M_1$, as defined in the main text. Here, we make explicit the mean values and uncertainties of $\Delta M$.

\subsubsection{Mean values of \texorpdfstring{$\Delta M$}{\it{DM}}}

In the rotating frame, the position component of $\langle \Delta M \rangle$ is: 
\begin{equation}\label{mean_Delta_x}
    \langle \Delta M_x \rangle = \left\langle \tilde{x}(\tint) - e^{-\kappa \tint/2}\tilde{x}(0) \right\rangle = 0,
\end{equation}
and the momentum component $\langle \Delta M_p \rangle$ is
\begin{equation}\label{mean_Delta_p}
    \langle \Delta M_p \rangle = \left\langle \tilde{p}(\tint) - e^{-\kappa \tint/2}\tilde{p}(0) \right\rangle = -\frac{4\hbar\lambda S_z}{\kappa \pi z_p}\left(1-e^{-\kappa \tint/2}\right),
\end{equation}
as given by the theory of Kalman filters and equation \eqref{formal_int_x_p} \footnote{By the definition of Kalman filters, we have that $\langle \hat{\tilde{x}}(t) \rangle = \langle \tilde{x}(t) \rangle = e^{-\kappa t/2}\tilde{x}(0) = e^{-\kappa t/2}\langle \hat{\tilde{x}}(0) \rangle $, where $\tilde{x}(0)$ is the initial state.}.
\subsubsection{Covariance of \texorpdfstring{$\Delta M$}{\it{DM}}}
The position variance is
\begin{equation}
    \left\langle \Delta M_x^2 \right\rangle = \left\langle \left(\hat{\tilde{x}}(\tint) - e^{-\kappa \tint/2}\hat{\tilde{x}}(0)\right)^2 \right\rangle = \left\langle \hat{\tilde{x}}(\tint)^2 \right\rangle - 2e^{-\kappa \tint/2} \left\langle \hat{\tilde{x}}(\tint) \hat{\tilde{x}}(0) \right\rangle + e^{-\kappa \tint} \left\langle \hat{\tilde{x}}(0))^2 \right\rangle.
\end{equation}
We employ equation \eqref{var_final_est_pos} to retrieve the first and last term. Since the measurement noise and diffusion are independent, we have that
\begin{equation}
    \left\langle \hat{\tilde{x}}(\tint) \hat{\tilde{x}}(0) \right\rangle = \left\langle \hat{\tilde{x}}(\tint) \right\rangle \left\langle \hat{\tilde{x}}(0) \right\rangle = e^{-\kappa \tint/2} \tilde{x}(0).
\end{equation}
The position variance is
\begin{equation}\label{delta_pos_var}
    \left\langle \Delta M_x^2 \right\rangle = \left\langle \left(\hat{\tilde{x}}(\tint) - e^{-\kappa \tint/2}\hat{\tilde{x}}(0)\right)^2 \right\rangle = \Delta m^2\left(1+e^{-\kappa \tint}\right) + n_{th}\left(1-e^{-\kappa \tint}\right) \equiv \sigma(\tint)^2,
\end{equation}
as defined in the main text. A similar calculation shows that the momentum variance is also equal to $\sigma(\tint)^2$.

\subsubsection{Minimum pulse sequence duration} 

Since $\sigma(\tint)$ is a very important quantity in our scheme, we analyze the relative sizes of the two sources of uncertainty (measurement noise and diffusion). Taking the ratio of the two variances at small $\kappa t$, and stipulating that the diffusion dominates the variance (i.e., $\Delta m^2 \ll \kappa n_{th} \tint$) we find that
\begin{equation}
    \frac{\Delta m^2}{n_{th} \kappa \tint} = \frac{\lambda_l}{\pi z_p} \sqrt{\frac{\kappa n_{th}}{R}} \frac{1}{\kappa n_{th}\tint} \ll 1.
\end{equation}
For this quantity to be much less than 1, we arrive at the inequality
\begin{equation}\label{condition_on_t}
    \tint \gg \frac{\lambda_l}{\pi z_p}\sqrt{\frac{1}{\kappa R n_{th}}}.
\end{equation}
For reasonable parameters, it is possible to have the condition \eqref{condition_on_t} to be satisfied within a typical interaction time. For example, for resonators with $z_p \sim 10^{-14}$m and at room temperature, a resonance frequency of $\omega_r\sim1$~MHz and quality factor of $Q\sim10^7$, as well as a 1~mW infrared scattered beam, we have $t \gg 0.2$~ms. However, for smaller $n_{th}$, higher $Q$, or short $\tint$, we may require an optical cavity to enhance the measurement efficiency.

\section{Uncertainty of a shot noise limited measurement}
In the following we provide an alternative derivation of the measurement uncertainty to provide some insights into how it depends on the measurement time. We consider a shot noise limited position measurement, where the imprecision power spectral density (PSD) is \if\suppl{0}\footnote{our definition of the power spectral density is $\langle x^2\rangle = \int_{-\infty}^\infty S_{xx}(\omega){d}\omega$.}\fi
\begin{equation}
S_{\rm imp} =  \frac{1}{\eta_{\rm geo}\eta_{\rm det}} \frac{\hbar c}{\pi k} \frac{1}{P}
\end{equation}
where $\eta_{\rm geo}$ is a geometric factor \footnote{For a dipole measured in back-reflection, this would be $\eta_{\rm geo} =8 (\frac{2}{5}+A^2)$, with $0.64 \leq A \leq 1$ and $\eta_{\rm det}$ the detection efficiency of the setup}, $k = 2\pi/\lambda_l$ is the wavevector, $c$ is the speed of light, $P$ is the scattered power and $\eta_{\det}$ is the detection efficiency.
Introducing the rate at which scattered photons are collected $R =\eta_{\rm geo}\eta_{\rm det} P / (\hbar c k)$, we can write the above as 
\begin{equation}
S_{\rm imp} =  \frac{1}{\pi k^2 R}
\end{equation}
Since the power imprecision PSD is flat, the integrated imprecision over a measurement of duration $\tau_m$ is simply given by $\langle x^2\rangle = (2\pi/\tau_m) S_{\rm imp} $.
Introducing the position uncertainty normalized by the zero point motion, we find
\begin{equation}\label{eq:dm_imp}
  \Delta m_{\rm imp}^2  = \frac{\langle x^2 \rangle}{2 z_p^2} = \left(\frac{\lambda_l}{2\pi z_p}\right)^2 \left[  R \tau_m\right]^{-1}.
\end{equation}
During the measurement, the position undergoes diffusion due to the interaction with the thermal environment. This changes the number of excitations by $\Delta m_{\rm th}^2 = \kappa n_{\rm th}\tau_m$, where $n_{\rm th} = k_B T / (\hbar \omega_r)$ is the thermal occupation and $\kappa$ the coupling or damping rate.
Hence, the total uncertainty after time $\tau_m$ is
\begin{equation}\label{eq:dm_tot}
  \Delta m_{\rm tot}^2 = \Delta m_{\rm imp}^2 + \Delta m_{\rm th}^2 = 
  \left(\frac{\lambda_l}{2\pi z_p}\right)^2 \left[  R \tau_m\right]^{-1} 
  + \kappa n_{\rm th}\tau_m
\end{equation}
The optimal time, that minimizes \eqref{eq:dm_tot} is given by 
\begin{equation}\label{eq:tau_opt}
  \tau_m^{\rm opt} = \left(\frac{\lambda_l}{2\pi z_p}\right) \left[  R n_{\rm th}\kappa \right]^{-1/2}
\end{equation}
when the imprecision due to diffusion equals the measurement imprecision. Therefore, the total uncertainty at the optimal measurement time is
\begin{equation}\label{eq:total_uncertainty}
   \Delta m_{\rm opt}^2  = 2\left(\frac{\lambda_l}{2\pi z_p}\right) \sqrt{\frac{\kappa n_{\rm th}}{ R}}
\end{equation}
Note that this result agrees with our previous derivation for the Kalman filter \eqref{Delta_m}. 
This is not surprising, since the Kalman filter is known to be optimal and allows to extract all the available information.

With the above, we can now also extend the discussion to the case of a strong measurement, where the measurement backaction dominates over the thermal diffusion.
Unlike thermal diffusion, which acts symmetrically on the position and momentum quadratures, the measurement backaction acts only on the position quadrature. 
Hence, for the momentum quadrature the scaling of the uncertainty with scattered photons continues to decrease with increasing laser power following \eqref{eq:total_uncertainty}.

In contrast, for the position quadrature we have to add the backaction term $\propto R$ to \eqref{eq:dm_tot}.
As a consequence, the uncertainty does not monotonlically decrease with laser power for the positon qudrature and instead there exist an optimum value given by the Heisenberg uncertainty principle.

\section{Scaling with cooperativity}
In this section we discuss the details of the scaling of the error with cooperativity. 
\subsection{Local scaling \texorpdfstring{$p(C)$}{\it{p(C)}}} 
Suppose we write the error as $\mathcal{E} = \mathcal{E}_0 C^{p(C)}$ where $p(C)$ is an exponent that depends on the cooperativity. We quantify the local scaling exponent $p(C)$ for $\alpha\rightarrow0$ to be
\begin{equation}
    p(C) = \frac{\text{d}\ln\mathcal{E}}{\text{d}\ln C} \approx  -\frac{\pi^2}{8}\frac{\left(16C/\pi^2-1\right)^{\pi^2/8C}\text{ln}\left(16C/\pi^2-1\right)}{C\left(\left(16C/\pi^2-1\right)^{\pi^2/8C}-1\right)}.
\end{equation}
This expression converges to -1 for $C\rightarrow\infty$.

\begin{figure}
\centering
\includegraphics[width=0.4\textwidth]{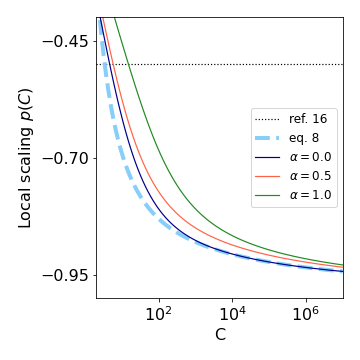}
\caption{Scaling of the error $p(C) \equiv \frac{\text{d}\ln\mathcal{E}}{\text{d}\ln C}$ as a function of cooperativity for various $\alpha$. The blue dashed curve is the scaling of the fidelity using the analytic approximation to the optimal time as defined in the main text.}
\label{fig:p_C_scaling}
\end{figure}

To derive this, we first define the expression $f(A)=(2A-1)^{1/A}$. We find that $\forall A>1$, $f(A)>1$, as the base exceeds 1. We further find for arbitrarily small $\epsilon>0$, $\exists A_0$, such that $(2A-1)<(1+\epsilon)^A$ $\forall A>A_0$ and thus $f(A)<1+\epsilon$. Consequently, $\lim_{A\rightarrow \infty} f(A) = 1^+$. More specifically, in the expansion $1/f(A)$
\begin{equation}
    1/f(A) =\exp\left(-\frac{\ln\left(2A-1\right)}{A}\right) \approx 1 - \frac{\ln\left(2A-1\right)}{A} \quad\text{for}~A\rightarrow \infty,
\end{equation}\label{expansion_inv_f(A)}
such that $\lim_{A\rightarrow \infty} 1/f(A) = 1^-$. 
Using the substitution $A \equiv 8C/\pi^2$ and $1/f(A)\equiv y$, the limiting local exponent is
\begin{equation}
    \lim_{C\rightarrow\infty} p(C) = \lim_{y\rightarrow1^-} \frac{\ln(y)}{1-y} = -1.
\end{equation}
See figure \ref{fig:p_C_scaling} for an illustration for the scaling $p(C)$ for various $\alpha$ threshold parameters and finite $C$.
\subsection{Scaling of error at large \texorpdfstring{$C$}{\it{C}}}
To show the $\mathcal{E} \sim \text{ln}C/C$ scaling, with $A$, $f(A)$ as defined above, we can express the error of the Bell state preparation as 
\begin{align}
    \mathcal{E} & = 1 - \frac{\left(2A-1\right)\left(1+\frac{1}{f(A)}\right)}{4A}\\
    & = \frac 1 2 - \frac 1 2 \frac{1}{f(A)} + \frac{1}{4A} + \frac{1}{4A f(A)}.\label{SI:scaling:e1}
\end{align}
Inserting the Taylor expansion of the expression in \eqref{expansion_inv_f(A)} into eq. \eqref{SI:scaling:e1}, we find the leading order 
\begin{align*}
    \mathcal{E} & = \frac{\ln({2A-1})+1}{2A} + \mathcal{O}\!\left(\frac{\ln A}{A^2}\right)\\
    & \approx \frac{\pi^2}{16}\frac{\ln C}{C}+\frac{\pi^2\left(1+\ln\frac{16}{\pi^2}\right)}{16}\frac{1}{C} \quad\text{for}~C\rightarrow \infty
\end{align*}
of the error of the Bell state preparation. 

\subsection{Review of deterministic hot gate - scaling of error with \texorpdfstring{$C$}{\it{C}}} 
Previously proposed hot gates \cite{schuetz_high-fidelity_2017, rabl_quantum_2010} require two conditions. One is that the entanglement rate is faster than the phonon induced spin decoherence 
\begin{equation}
\begin{split}
    \lambda^2/\omega_r > (\lambda / \omega_r)^2 \kappa n_{th}
    \rightarrow Q > n_{th},
\end{split}
\end{equation}
and the other is that the entanglement rate is faster than the intrinsic spin decoherence,
\begin{equation}
    \lambda^2 / \omega_r > 1/T_2 = \Gamma.
\end{equation}
Combining both of these errors, following \cite{schuetz_high-fidelity_2017}, we have that the total error is
\begin{equation}
    \mathcal{E} \sim \alpha_k \kappa n_{th}/\omega_r + \alpha_T \Gamma \omega_r/\lambda^2.
\end{equation}
As discussed in \cite{schuetz_high-fidelity_2017}, this is minimized at an \emph{optimal resonator frequency} $\omega_{r, opt}$, that is a function of the temperature $T$ and $Q$ factor of the resonator.  The pre-factors $\alpha_k$ and $\alpha_T$ are found numerically in \cite{schuetz_high-fidelity_2017} to be $4$ and $0.1$, respectively. We assume that $\lambda$ is independent of the resonator frequency: the gradient is independent of frequency, and the zero point motion scales as $\sim 1/\sqrt{m \omega_r}$. When scaling the resonator frequency of a nanobeam resonator, we generally change the resonator length: this results in a frequency change of $\sim 1/L$ and a mass change $\sim L$, such that the zero point motion is also independent of the changing frequency. The optimal frequency as found in \cite{schuetz_high-fidelity_2017} is:
\begin{equation}
    \omega_{r, opt} = \lambda \sqrt{\frac{\kappa n_{th}}{\Gamma}\frac{\alpha_k}{\alpha_T}},
\end{equation}
such that the error at the optimal frequency as found in \cite{schuetz_high-fidelity_2017} is
\begin{equation}
    \mathcal{E} \sim 1.2/\sqrt{C}.
\end{equation}

\section{Application: CNOT gate}

As discussed in the main text, one application of the entanglement scheme is to use the generated entangled state between the two spins to teleport a CNOT gate between other nearby, coupled spins. In this section, we elaborate on other possible sources of error accumulated over the gate teleportation circuit specific to NV centers.

\subsection{Charge state initialization} The charge state initialization error of NV- electronic spins $\mathcal{E}_{charge}$ can be as high as 25\% but can be decreased to 0.5\% using real-time feedback without any additional delay \cite{hopper_real-time_2020}, and can be further improved with doping \cite{doi_deterministic_2014, doi_pure_2016} or post-selecting for the negative charge state \cite{waldherr_quantum_2014}. 

\subsection{NV-nuclear spin CNOT errors} 

The two-qubit gate error $\mathcal{E}_{CNOT}$ of optimized sequences is expected to be between $\sim 10^{-5}$ and $\sim 10^{-1}$, depending on the particular values of the NV-nuclear hyperfine coupling and the pulse sequence \cite{dong_precise_2020, chou_optimal_2015} while experimentally, error rates of less than $10^{-2}$ have been demonstrated in NV-$^{13}$C systems \cite{bradley_ten-qubit_2019} at cryogenic temperatures and several percent at room temperature \cite{van_der_sar_decoherence-protected_2012, waldherr_quantum_2014}. 

\subsection{Mechanical readout of the NV spin state}\label{subsection:mech_readout}
As the readout of the spin state does not require to preserve the coherence of the state, it will not be limited by the spin coherence time $1/\Gamma$ but by the spin life time $T_1=1/\Gamma_1$. We find that the expectation value for the displacement in case of a finite spin life-time after a pulse sequence as described in the main text is applied to spin $i$ for time $t_{RO}$ is
\begin{equation}
    \left\langle\mu\right\rangle = \frac{2\sqrt{2}\lambda \sigma_z^{(i)}}{\pi(\kappa/2-\Gamma_1)}\left(e^{-\Gamma_1 t_{RO}}-e^{-\kappa t_{RO}/2}\right).
\end{equation}
For $t_{RO}\ll 1/\kappa, 1/\Gamma_1$, this simplifies to $2\sqrt{2}\lambda t_{RO} \sigma_z^{(i)}/\pi$, as described in the main text. As $t_{RO}$ is typically very short compared to $T_1$ in order to preserve the nuclear spin coherence, we take the limit $T_1\rightarrow\infty$ for calculating the mechanical readout fidelity. We proceed in close analogy to the entanglement protocol, i.e.\ we determine the resonator state, then apply the pulse sequence to spin $i$, and we measure the resonator again. A positive (negative) displacement of the resonator indicates $\sigma_z^{(i)}=+1$ ($-1$). The fidelity of this assignment can be estimated as
$\mathcal{F}_{RO}=1 - \frac 1 2 \text{Erfc}\left(\left\langle\mu\right\rangle/\sqrt{2}\sigma\right)$. 
In the limit of $\Delta m^2\ll \kappa n_{th} t_{RO}$ and $t_{RO}\ll 1/\kappa, 1/\Gamma_1$, we find the associated readout error $\mathcal{E}_{RO}= \frac 1 2 \text{Erfc}\left( \frac{\lambda}{\pi}\sqrt{\frac{4 t_{RO}}{\kappa n_{th}}} \right)$. This mechanical spin readout can be used for feedback based initialization, eliminating the need for a spin photon interface for this entanglement protocol. This furthermore eliminates the need for continuous charge state control.

\subsection{Optical readout and initialization of the NV spin state}
While implementing a repetitive readout scheme using a second, nearby nuclear spin and a quantum nondemolition hyperfine coupling \cite{jiang_repetitive_2009}, the readout error $\mathcal{E}_{RO}$ depends on the hyperfine coupling strengths as well as magnetic field and alignment \cite{hopper_spin_2018}, but has been measured to be as low as $\sim4\cdot 10^{-2}$ \cite{dreau_single-shot_2013, jiang_repetitive_2009}, and can approach $5\cdot 10^{-3}$ with machine learning classification \cite{liu_repetitive_2020}.

However, optical illumination of the NV induces decoherence on the nuclear spin leading to significant errors during repetitive readout and optical illumination which can be significant \cite{reiserer_robust_2016}.

\subsection{Hyperfine interaction}
Another contribution to $\mathcal{E}_{int}$ arises from the NV-nuclear spin hyperfine interaction during the pulse sequence. In a dynamical decoupling sequence, the hyperfine coupling component orthogonal to the bias field $A_{\perp}$ between the electron and any nearby nucleus can lead to conditional spin rotations. By using the $\ket{\pm1}$ NV spin states as qubit basis, and choosing a bias field for which the nuclear Lamor frequency $\omega_{L,n}$ strongly exceeds $A_{\perp}, A_{||}$, the resonances corresponding to conditional spin rotations of all nuclear spins with the same gyromagnetic ratio (e.g.\ all $^{13}C$ atoms) stay closely aligned \cite{nguyen_integrated_2019}. By ensuring that $\omega_{L,n}$ and $\omega_r$ are strongly detuned, the unwanted hyperfine interaction can be minimized. Residual dephasing can be compensated by inverting the nuclear spins at time $\tint/2$. 

\subsection{Pulse errors}

The UR20 sequence reported in \cite{genov_arbitrarily_2017} has error $\sim {\epsilon}^{10}$ for 20 pulses, where $\epsilon$ is the single pulse error. If we assume $\epsilon \sim 10^{-3}$, corresponding to state-of-the-art single qubit gates on NV centers \cite{rong_experimental_2015}, the contribution of pulse errors to the total error budget are expected to be negligible. 

\section{Effects of finite \texorpdfstring{$\kappa \tint$}{kt}}

\begin{figure}
\centering
\includegraphics[width=0.5\textwidth]{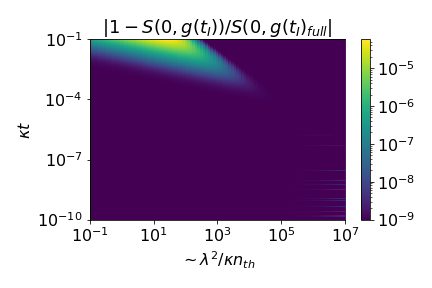}
\caption{The success probability $S(0, g(t_I)_{full})$ compared to $S(0, g(t_I)$ as given in the main text. We add $10^{-9}$ to all results to avoid numerical errors as the result approaches zero.}
\label{fig:finite_kt}
\end{figure}

Since we expand $g(\tint)$ for small $\kappa \tint$ and then insert the result into the equation for the success probability $S$ given in the main text, we here confirm that the resulting $S$ is the same as if we did not expand $g(\tint)$. We find that the success probability $S(0, g(\tint))$ using the first order expansion of $g$ as in the main text is the same as the success probability $S(0, g(\tint)_{full})$ without any $g$ expansion, to a few parts in $10^5$ or below in our region of interest. See figure \ref{fig:finite_kt}.

\appendix
\section{Table of variables and definitions}

\begin{center}
  \begin{tabular}{ | l | c | r |}
    \hline
    \textbf{Name} & \textbf{Description} & \textbf{Units} \\ \hline
    $\lambda$ & spin-phonon coupling & 2$\pi$ Hz \\ \hline
    $z_p$ & resonator zero-point motion & m \\ \hline
    $S_z$ & total spin angular momentum in z & none \\ \hline
    $\omega_s$ & spin resonance frequency & 2$\pi$ Hz \\ \hline
    $a (a^{\dagger})$ & resonator bosonic operators & none \\ \hline
    $\omega_r$ & resonator frequency & 2$\pi$ Hz \\ \hline
    $k_B T$ & 1/thermodynamic $\beta$ & J \\ \hline 
    $n_{th}$ & thermal occupation number & none \\ \hline
    $K$ & Kalman gain & matrix, multiple \\ \hline
    $\lambda_l$ & laser wavelength & m \\ \hline 
    $\kappa $ & resonator bandwidth & 2$\pi$ Hz \\ \hline 
    $ Q = \omega_r / \kappa $ & resonator Q factor & none \\ \hline 
    $D$ & resonator diffusion constant & kg$^2$ m$^2$/s$^3$ \\ \hline
    $m$ & resonator mass & kg \\ \hline
    $\hbar$ & Plank's constant/2$\pi$ & J s \\ \hline
    $P_{ss}$ & steady-state covariance & matrix, multiple \\ \hline
    $\Delta m^2$ & variance in momentum (and position) from $P_{ss}$ & none \\ \hline
    $P$ & interferometer scattered laser power & W \\ \hline
    $E$ & energy of an interferometer photon & J \\ \hline
    $R \equiv P/E$ & Scattered photon flux rate & 1/s \\ \hline
    $x(t)$ & resonator state position, lab frame & m \\ \hline
    $p(t)$ & resonator state momentum, lab frame & kg m/s \\ \hline    
    $\tilde{x}(t)$ & resonator state position, rotating frame & m \\ \hline
    $\tilde{p}(t)$ & resonator state momentum, rotating frame & kg m/s \\ \hline
    $\hat{x}(t)$ & estimated position, lab frame & kg m/s \\ \hline
    $\hat{p}(t)$ & estimated momentum, lab frame & kg m/s \\ \hline
    $\tilde{\hat{x}}(t)$ & estimated position, rotating frame & kg m/s \\ \hline
    $\tilde{\hat{p}}(t)$ & estimated momentum, rotating frame & kg m/s \\ \hline
    \hline
    
    \hline
  \end{tabular}
\end{center}
\if\suppl0
\bibliographystyle{unsrt}

\bibliography{dfs_entangle} 
\fi

\end{document}

\else 

\end{document}